\documentclass[10pt]{article}
\usepackage{xspace}
\usepackage{amssymb}
\usepackage{subfigure}
\usepackage{graphicx}

\title{Verifying Sequential Consistency on Shared-Memory Multiprocessors by Model Checking}
\author{
Shaz Qadeer \\ 
Compaq Systems Research Center \\ 
130 Lytton Ave \\ 
Palo Alto, CA 94301 \\
{\tt shaz.qadeer@compaq.com}
}
\date{}

\begin{document}
\maketitle

\newcommand{\fact}[1]{\ensuremath{{#1}!}}
\newcommand{\lang}[1]{\ensuremath{\overline{#1}}}
\newcommand{\cons}{\mathit{Constrain}}
\newcommand{\checker}{\mathit{Check}}
\newcommand{\upto}{\mathit{upto}}
\newcommand{\trace}[1]{\ensuremath{\overline{#1}}}
\newcommand{\store}{{\it st}}
\newcommand{\zug}[1]{\mbox {\ensuremath{\langle #1 \rangle }}}
\newcommand{\set}[1]{\ensuremath{\{#1\}}}
\newcommand{\ordera}{\ensuremath{<}}
\newcommand{\orderb}{\ensuremath{\prec}}
\newcommand{\minimum}[1]{\mbox {\ensuremath{min({#1})}} }
\newcommand{\maximum}[1]{\mbox {\ensuremath{max({#1})}} }
\newcommand{\msystem}{S}
\newcommand{\op}{{\it op}}
\newcommand{\proc}{{\it proc}}
\newcommand{\loc}{{\it loc}}
\newcommand{\data}{{\it data}}
\newcommand{\proj}{\ensuremath{\rho}}
\newcommand{\nat}{{\mathbb N}}
\newcommand{\natzero}{{\mathbb W}}
\newcommand{\dataset}{\nat}
\newcommand{\memevent}{{\it E}}
\newcommand{\intevent}{{\it F}}
\newcommand{\rdevent}{\ensuremath{\memevent^r}}
\newcommand{\wrtevent}{\ensuremath{\memevent^w}}
\newcommand{\event}{\ensuremath{E^a}}
\newcommand{\rd}{{\it R}}
\newcommand{\wrt}{{\it W}}
\newcommand{\mb}{{\it MB}}
\newcommand{\witness}{\ensuremath{\Omega}}
\newcommand{\procorder}[3]{\ensuremath{{#1}({#2},{#3})}}
\newcommand{\locorder}[3]{\ensuremath{{#1}({#2},{#3})}}
\newcommand{\expand}[1]{\ensuremath{{#1}^e}}
\newcommand{\domain}{{\it dom}}
\newcommand{\dreduce}{\ensuremath{\Gamma}}
\newcommand{\order}{{\it order}}
\newcommand{\compose}{\ensuremath{\circ}}
\newcommand{\err}{{\it err}}
\newcommand{\msc}{\ensuremath{M}}
\newcommand{\malpha}{\ensuremath{\mmodel^{\alpha}}}
\newcommand{\StatesSC}{\ensuremath{\Sigma^{sc}}}
\newcommand{\state}[2]{\ensuremath{[\![#1]\!](#2)}}

\newcommand{\LET}{{\it let}\xspace}
\newcommand{\IN}{{\it in}\xspace}
\newcommand{\IF}{{\it if}\xspace}
\newcommand{\OF}{{\it of}\xspace}
\newcommand{\ELSE}{{\it else}\xspace}
\newcommand{\THEN}{{\it then}\xspace}
\newcommand{\FORALL}{{\it forall}\xspace}
\newcommand{\typedef}{{\it typedef}\xspace}
\newcommand{\Msg}{{\it Msg}\xspace}
\newcommand{\Entry}{{\it CacheEntry}\xspace}
\newcommand{\Queue}{{\it Queue}\xspace}
\newcommand{\record}{{\it record}\xspace}
\newcommand{\ARRAY}{{\it array}\xspace}
\newcommand{\Proc}{{\it Proc}\xspace}
\newcommand{\Addr}{{\it Addr}\xspace}
\newcommand{\Data}{{\it Data}\xspace}
\newcommand{\Index}{{\it Index}\xspace}
\newcommand{\INVAL}{{\it INVAL}\xspace}
\newcommand{\SHDACK}{{\it ACKS}\xspace}
\newcommand{\EXCACK}{{\it ACKX}\xspace}
\newcommand{\INV}{{\it INV}\xspace}
\newcommand{\SHD}{{\it SHD}\xspace}
\newcommand{\EXC}{{\it EXC}\xspace}
\newcommand{\head}{{\it head}\xspace}
\newcommand{\tail}{{\it tail}\xspace}
\newcommand{\cache}{{\it cache}\xspace}
\newcommand{\inQ}{{\it inQ}\xspace}
\newcommand{\owner}{{\it owner}\xspace}
\newcommand{\RD}{{\it RD}\xspace}
\newcommand{\WR}{{\it WR}\xspace}
\newcommand{\EXCRSP}{{\it ACKX}\xspace}
\newcommand{\SHDRSP}{{\it ACKS}\xspace}
\newcommand{\UPDATE}{{\it UPD}\xspace}
\newcommand{\full}{{\it full}\xspace}
\newcommand{\EMPTY}{{\it isEmpty}\xspace}
\newcommand{\APPEND}{{\it append}\xspace}
\newcommand{\HEAD}{{\it head}\xspace}
\newcommand{\TAIL}{{\it tail}\xspace}
\newcommand{\msg}{{\it msg}\xspace}

\newtheorem{definition}{Definition}
\newtheorem{theorem}{Theorem}[section]
\newtheorem{lemma}[theorem]{Lemma}
\newtheorem{claim}[theorem]{Claim}
\newtheorem{corollary}[theorem]{Corollary}
\newtheorem{proposition}[theorem]{Proposition}
\newtheorem{assumption}{Assumption}
\newtheorem{requirement}{Requirement}
\def\EndProof{\hfill \quad \vrule width 1ex height 1ex depth 0pt  }
\newenvironment{proof}{\noindent{\bf Proof}:}{\EndProof}
\newenvironment{example}{{\bf Example}.}{\EndProof}


\begin{abstract}

The memory model of a shared-memory multiprocessor is a contract
between the designer and programmer of the multiprocessor.  
The sequential consistency memory model specifies a total order
among the memory (read and write) events performed at each processor. 
A trace of a memory system satisfies sequential consistency if there exists a
total order of all memory
events in the trace that is both consistent with the total order at each processor
and has the property that every read event to a location 
returns the value of the last write to that location.

Descriptions of shared-memory systems are typically parameterized by the
number of processors, the number of memory locations, and the number
of data values.
It has been shown that even for finite
parameter values, verifying sequential consistency on general shared-memory
systems is undecidable.  
We observe that, in practice, shared-memory systems satisfy the properties of 
causality and data independence.
Causality is the property that values of read events flow from values
of write events. 
Data independence is the property that all traces can be generated by
renaming data values from traces where the written values are distinct
from each other. 
If a causal and data independent system also has the property that  
the logical order of write events to each location 
is identical to their temporal order, then sequential consistency can be 
verified algorithmically.
Specifically, we present a model checking algorithm to verify sequential consistency 
on such systems 
for a finite number of processors and memory locations and an arbitrary 
number of data values.
\end{abstract}

\section{Introduction}
Shared-memory multiprocessors are very complex computer systems.  
Multithreaded programs running on shared-memory
multiprocessors use an abstract view of the shared memory that is specified
by a memory model.  
Examples of memory models for multiprocessors include 
sequential consistency \cite{Lamport79}, partial store ordering
\cite{sparc-manual}, and the Alpha memory model \cite{alpha-manual}.
The implementation of the memory model, achieved by a protocol running 
either in hardware or software, is one of the most complex aspects of 
multiprocessor design.
These protocols are commonly referred to as cache-coherence protocols.
Since parallel programs running on such systems rely on the memory
model for their correctness, it is important to implement the
protocols correctly. 
However, since efficiency is important for the commercial viability
of these systems,
the protocols are heavily optimized, making them prone to design errors.
Formal verification of cache-coherence protocols can detect these
errors effectively.

Descriptions of cache-coherence protocols are typically parameterized
by the number of processors, the number of memory locations, and the
number of data values.
Verifying parameterized systems for arbitrary values of
these parameters is undecidable for nontrivial systems.
Interactive theorem proving is one approach to parameterized
verification. 
This approach is not automated and is typically expensive in terms of
the required human effort. 
Another approach is to model check a parameterized system for small
values of the parameters.
This is a good {\em debugging\/} technique that can find a number of
errors prior to the more time-consuming effort of verification for
arbitrary parameter values.
In this paper, we present an automatic method based on model checking to
verify that a cache-coherence protocol with fixed parameter values
is correct with respect to the sequential consistency memory model.

The sequential consistency memory model \cite{Lamport79} specifies a
{\em total order\/} among 
the memory events (reads and writes) performed locally at each
processor.
This total order at a processor is the order in which memory events
occur at that processor.
A trace of a memory system satisfies sequential consistency 
if there exists a total order of all memory events that 
is both consistent with the local total order at each processor, 
and has the property that every read to a location returns the latest
(according to the total order) value written to that location. 
Surprisingly, verifying sequential consistency, even for fixed
parameter values, is undecidable \cite{AMP96b}. 
Intuitively, this is because the witness total order
could be quite different from the global temporal order of events 
for some systems.
An event might need to be logically ordered after an event that 
occurs much later in a run.
Hence any algorithm needs to keep track of a potentially unbounded 
history of a run.

In this paper, we consider the problem of verifying that a
shared-memory system $\msystem(n,m,v)$ with $n$ processors, $m$
locations and $v$ data values is sequentially consistent.
We present a method that can check sequential consistency for any 
fixed $n$ and $m$ and for arbitrary $v$.
The correctness of our method depends on two assumptions
---causality and data independence.
The property of {\em causality\/} arises from the observation that 
protocols do not conjure up data values; data is injected into the
system by the initial values stored in the memory and by the writes
performed by the processors.
Therefore every read operation $r$ to location $l$ is associated
with either the initial value of $l$ or some 
write operation $w$ to $l$ that wrote the value read by $r$.
The property of {\em data independence\/} arises from the observation that 
protocols do not examine data values; they just forward the
data from one component of the system (cache or memory) to another.
Since protocol behavior is not affected by the data values, we can
restrict our attention, without loss of generality, to unambiguous
runs in which the 
written data values to a location are distinct from each other and from 
the initial value.
We have observed that these two assumptions are true of shared-memory
systems that occur in practice \cite{LLG90,KOH94,BDH99,BGM00}. 

For a causal and unambiguous run, we can 
deduce the association between a read and the associated write just
by looking at their data values.
This leads to a vast simplification in the task of specifying the
witness total order for sequential consistency.
It suffices to specify for each location, a total order on the writes
to that location.
By virtue of the association of write events and read events, the 
total order on the write events can be extended to a partial order 
on all memory events (both reads and writes) to that location. 
If a read event~$r$ reads the value written by the write event~$w$,
the partial order puts~$r$ {\em after\/}~$w$ and all write events
preceding~$w$, and {\em before\/} all write events succeeding~$w$.
As described before, sequential consistency specifies 
a total order on the memory events for each processor.
Thus, there are $n$ total orders, one for each processor,
and $m$ partial orders, one for each location, imposed on the graph of
memory events of a run.
A necessary and sufficient condition for the run to be
sequentially consistent is that this graph is acyclic.
We further show that existence of a cycle in this graph implies the existence of a 
{\em nice\/} cycle in which no two processor edges (imposed by the memory model) 
are for the same processor
and no two location edges (imposed by the write order) are for the same location.  
This implies that a nice cycle can have at most $2 \times \minimum{\{n,m\}}$ edges; 
we call a nice cycle with $2 \times k$ edges a $k$-nice cycle.
Further if the memory system is symmetric with respect to processor
and location ids, then processor and location edges occur in a certain
canonical order in the nice cycle.
These two observations drastically reduce the number of cycles for any search.

We finally argue that a number of causal and data independent shared-memory systems
occurring in practice also have the property that the witness write order at each 
location is simply the temporal order of the write events.
In other words, a write event $w$ is ordered before $w'$ if $w$ occurs
before $w'$.
We call this a simple write order, and it is in fact
the correct witness for a number of shared-memory systems.
For cache-based shared-memory systems, the intuitive explanation is
that at any time there is at most one cache with
write privilege to a location.
The write privilege moves from one cache to another with time.
Hence, the logical timestamps \cite{Lamport78} of the writes to a
location order them exactly according to their global temporal order.
We show that the proof that a simple write order is a correct witness for a
memory system can be performed by model checking
\cite{ClarkeEmerson81, QueilleSifakis81}.
Specifically, the proof for the memory system $\msystem(n,m,v)$ for
fixed $n$ and $m$ and arbitrary $v$ is broken into 
$\minimum{\{n,m\}}$ model checking lemmas, where the $k$-th lemma checks for the 
existence of canonical $k$-nice cycles.


The rest of the paper is organized as follows. 
Sections~\ref{sec:msystem} and~\ref{sec:causal-data-ind} formalize
shared-memory systems and our assumptions of causality and data
independence about them.
Section~\ref{sec:mmodel} defines the sequential consistency memory
model.
Section~\ref{sec:witness} defines the notions of a witness and
a constraint graph for an unambiguous and causal run.
Section~\ref{sec:cycle} and~~\ref{sec:symmetry} show that it is
sufficient to search for canonical nice cycles in the constraint graph.
Section~\ref{sec:model-check} shows how to use model checking to
detect canonical nice cycles in the constraint graphs of the runs of a
memory system.
Finally, we discuss related work in Section~\ref{sec:rel-work} and
conclude in Section~\ref{sec:conclusions}.

\section{Shared-memory systems}
\label{sec:msystem}

\begin{figure}
\begin{tabbing}
$\zug{\SHDRSP,i,j}~~$ \= xx \= xx \= xx \= xx \= xx \= xx \= xx \= \kill 
$\typedef\ \Msg\ \{m : \{\SHDACK, \EXCACK\}, a : \nat_m, d : \natzero_v\} \cup \{m : \{\INVAL\}, a : \nat_m\};$ \\
$\typedef\ \Entry\ \{d : \natzero_v, s : \{\INV, \SHD, \EXC\}\};$ \\
$\cache: \ARRAY\ \nat_n\ \OF\ \ARRAY\ \nat_m\ \OF\ \Entry;$ \\
$\inQ: \ARRAY\ \nat_n\ \OF\ \Queue(\Msg);$ \\
$\owner: \ARRAY\ \nat_m\ \OF\ \natzero_n;$ \\
\\
Initial predicate\\
$\forall i \in \nat_n, j \in \nat_m: (cache[i][j] = \zug{0,\SHD} \wedge
\inQ[i].\EMPTY \wedge owner[j] \neq 0)$ \\
Events\\
$\zug{\rd,i,j,k}$ \> $\cache[i][j].s \neq \INV \wedge \cache[i][j].d = k
\rightarrow$ \\
$\zug{\wrt,i,j,k}$ \> $\cache[i][j].s = \EXC \rightarrow$ \\
\>\> $\cache[i][j].d := k$ \\
$\zug{\EXCRSP,i,j}$ \> $\cache[i][j].s \neq \EXC \wedge \owner[j] \neq 0 \rightarrow$ \\
\>\>		   $\IF\ owner[j] \neq i\ \THEN\ \cache[\owner[j]][j].s := \INV;$ \\
\>\>	           $\owner[j] := 0;$ \\
\>\>		   for each $(p \in \nat_n)$ \\
\>\>\>	           $\IF\ (p = i)\ \THEN$ \\
\>\>\>\>              $\inQ[p] := \APPEND(\inQ[p],\zug{\EXCACK, j, \cache[\owner[j]][j].d})$ \\
\>\>\>		   $\ELSE\ \IF\ (p \neq \owner[j] \wedge
\cache[p][j].s \neq \INV)\ \THEN$ \\ 
\>\>\>\>	 	   $\inQ[p] := \APPEND(\inQ[p],\zug{\INVAL, j})$ \\
$\zug{\SHDRSP,i,j}$ \> $\cache[i][j].s = INV \wedge \owner[j] \neq 0 \rightarrow$ \\
\>\>			$\cache[\owner[j]][j].s := \SHD;$ \\
\>\>                    $owner[j] := 0;$ \\
\>\>			$\inQ[i] := \APPEND(inQ[i],\zug{\SHDACK, j,
\cache[\owner[j]][j].d});$ \\
$\zug{\UPDATE,i}$   \> $\neg \EMPTY(\inQ[i]) \rightarrow$ \\ 
\>\>			$\LET\ msg = \HEAD(\inQ[i])\ \IN$ \\
\>\>\>   	 	  $\IF\ (\msg.m = \INVAL)\ \THEN$ \\
\>\>\>\>                       $\cache[i][\msg.a].s := \INV$ \\
\>\>\>			  $\ELSE\ \IF\ (\msg.m = \SHDACK)\ \THEN\ \{$ \\ 
\>\>\>\>                       $\cache[i][\msg.a] := \zug{\SHD,\msg.d};$ \\
\>\>\>\>                       $\owner[\msg.a] := i$ \\
\>\>\>			  $\}\ \ELSE\ \{$ \\
\>\>\>\>                       $\cache[i][\msg.a] := \zug{\EXC,\msg.d};$ \\
\>\>\>\>                       $\owner[\msg.a] := i$ \\
\>\>\>                    $\}$ \\
\>\>                   $\inQ[i] := \TAIL(inQ[i])$
\end{tabbing}
\caption{Example of memory system}
\label{fig:example}
\end{figure}

Let $\nat$ denote the set of positive integers and $\natzero$
denote the set of non-negative integers.
For any $n \geq 1$, let $\nat_n$ denote the set of positive integers
up to $n$ and $\natzero_n$ denote the set of non-negative integers up
to $n$.

A memory system is parameterized by the number of processors, the
number of memory locations, and the number of data values.
The value $0$ is used to model the initial value of all memory
locations.
In a memory system with $n$ processors, $m$ memory locations, and $v$
data values, read and write events denoted by $\rd$ and $\wrt$ can
occur at any processor in $\nat_n$, to any 
location in $\nat_m$, and have any data value in $\natzero_v$ (the 
data values in $\nat_v$ together with the initial value $0$).
Formally, we define the following sets of events parameterized by the
number of processors $n$, the number of locations $m$, and the number
of data values $v$, where $n,m,v \geq 1$. 
\begin{enumerate}
\item
$\rdevent(n,m,v) = \{\rd\} \times \nat_n \times \nat_m \times
\natzero_v$ is the set of {\em read events\/}.
\item
$\wrtevent(n,m,v) = \{\wrt\} \times \nat_n \times \nat_m \times
\natzero_v$ is the set of {\em write events\/}.
\item
$\memevent(n,m,v) = \rdevent(n,m,v) \cup \wrtevent(n,m,v)$ is the 
set of {\em memory events\/}.
\item
$\event(n,m,v) \supseteq \memevent(n,m,v)$ is the set of 
{\em all events\/}.
\item
$\event(n,m,v) \setminus \memevent(n,m,v)$ is the set of 
{\em internal events\/}.
\end{enumerate}
The set of all finite sequences of events in $\event(n,m,v)$ is
denoted by $\event(n,m,v)^*$.
A {\em memory system\/} $\msystem(n,m,v)$ is a regular subset of 
$\event(n,m,v)^*$.
A sequence $\sigma \in \msystem(n,m,v)$ is said to be a {\em run\/}.
We denote by $\msystem(n,m)$ the union $\bigcup_{v \geq 1} \msystem(n,m,v)$.

Consider any $\sigma \in \event(n,m,v)^*$.
We denote the length of $\sigma$ by $|\sigma|$ and write $\sigma(i)$ for
the $i$-th element of the sequence.
The set of indices of the memory events in $\sigma$ is denoted by 
$\domain(\sigma) = \{1 \leq k \leq |\sigma|~|~\sigma(k) \in
\memevent(n,m,v)\}$.
For every memory event $e = \zug{a,b,c,d} \in \memevent(n,m,v)$, we define
$\op(e) = a$, $\proc(e) = b$, $\loc(e) = c$, and $\data(e) = d$.
The set of memory events by processor $i$ 
for all~$1 \leq i \leq n$ is denoted by 
$P(\sigma,i) = \{k \in \domain(\sigma)~ |~ \proc(\sigma(k)) = i\}$.
The set of memory events to location $i$ 
for all~$1 \leq i \leq m$ is denoted by 
$L(\sigma,i) = \{k \in \domain(\sigma)~ |~ \loc(\sigma(k)) = i\}$.
For all~$1 \leq i \leq m$,
the set of write events to location $i$ is denoted by 
$L^w(\sigma,i) = \{k \in L(\sigma,i)~ |~ \op(\sigma(k)) = \wrt\}$, and 
the set of read events to location $i$ is denoted by 
$L^r(\sigma,i) = \{k \in L(\sigma,i)~ |~ \op(\sigma(k)) = \rd\}$.

The subsequence obtained by projecting $\sigma$ onto $\domain(\sigma)$ is
denoted by $\trace{\sigma}$.
If $\sigma \in \msystem(n,m,v)$, the
sequence $\trace{\sigma}$ is a {\em trace\/} of $\msystem(n,m,v)$.
Similarly, if $\sigma \in \msystem(n,m)$, the sequence
$\trace{\sigma}$ is a {\em trace\/} of $\msystem(n,m)$.

\begin{example} 
Consider the memory system in Figure~\ref{fig:example}.
It is a highly simplified model of the protocol used to maintain cache
coherence within a single node in the Piranha chip multiprocessor system
\cite{BGM00}.
The system has three variables ---$\cache$, $\inQ$ and $\owner$--- 
and five events ---the memory events $\{\rd,\wrt\}$ and the internal 
events $\{\EXCRSP,\SHDRSP,\UPDATE\}$.
The variables $inQ$ and $\owner$ need some explanation.  
For each processor $i$, there is an input queue $\inQ[i]$ where
incoming messages are put. 
The type of $\inQ[i]$ is $\Queue$.
The operations $\EMPTY$, $\HEAD$ and $\TAIL$ are defined on $\Queue$,
and the operation $\APPEND$ is defined on $\Queue \times \Msg$.
They have the obvious meanings and their definitions have been omitted
in the figure. 
For each memory location $j$, either $\owner[j] = 0$ or $\owner[j]$
contains the index of a processor.
Each event is associated with a guarded command.
The memory events $\rd$ and $\wrt$ are parameterized by
three parameters ---processor $i$, location $j$ and data value $k$. 
The internal events $\EXCRSP$ and $\SHDRSP$ are parameterized by
two parameters ---processor $i$ and location $j$.
The internal event $\UPDATE$ is parameterized by processor $i$. 
A {\em state\/} is a valuation to the variables.
An {\em initial\/} state is a state that satisfies the initial predicate.
An event is {\em enabled\/} in a state 
if the guard of its guarded command is true in the state.
The variables are initialized to an initial state and updated by
nondeterministically choosing an 
enabled event and executing the guarded command corresponding to it. 
A run of the system is any finite sequence of events that can be executed
starting from some initial state.

A processor $i$ can perform a read to location $j$ if
$\cache[i][j].s \in \{\SHD,\EXC\}$, otherwise it requests $\owner[j]$
for shared access to location $j$.
The processor $\owner[j]$ is the last one to have received shared or
exclusive access to location $j$.
The request by $i$ has been abstracted away but the response of $\owner[j]$
is modeled by the action $\SHDRSP[i][j]$, which sends a $\SHDACK$ message
containing the data in location $j$ to $i$ and temporarily sets $\owner[j]$ to $0$.
Similarly, processor $i$ can perform a write to location $j$ if 
$\cache[i][j].s = \EXC$, otherwise it requests $\owner[j]$
for exclusive access to location $j$.
The processor $\owner[j]$ responds by sending a $\EXCACK$ message to
$i$ and $\INVAL$ messages to all other processors that have a valid
copy of location $j$.
$owner[j]$ is set to $i$ when processor $i$ reads the $\SHDACK$ or
$\EXCACK$ message from $\inQ[i]$ in the event $\UPDATE[i]$.
Note that new requests for $j$ are blocked while $\owner[j] = 0$.
A processor $i$ that receives an $\INVAL$ message for location $j$ sets
$\cache[i][j].s$ to $\INV$.
\end{example}

\section{Causality and data independence}
\label{sec:causal-data-ind}
In this section, we will state our main assumptions on memory
systems ---causality and data independence.

We assume that the runs of memory systems are {\em causal}.
That is, every read event to location $m$ ``reads'' either the initial value of 
$m$ or the value ``written'' to $m$ by some write event.
We believe that this assumption is reasonable because 
memory systems do not conjure up data values; they just move around
data values that were introduced by initial values or write events.
We state the causality assumption formally as follows.

\begin{assumption}[Causality]
\label{ass:causality}
For all $n,m,v \geq 1$, 
for all traces $\tau$ of $\msystem(n,m,v)$, 
and for all locations~$1 \leq i \leq m$,
if $x \in L^r(\tau,i)$, then either $\data(\tau(x)) = 0$ or 
there is $y \in L^w(\tau,i)$ such that 
$\data(\tau(x)) = \data(\tau(y))$.
\end{assumption}

The memory system in Figure~\ref{fig:example} is causal.
Only the write event $\wrt$ introduces a fresh data value in
the system by updating the cache; 
the internal events $\SHDACK$, $\EXCACK$ and $\UPDATE$ move data 
around and the read event $\rd$ reads the data present in the cache.
Therefore, the data value of a read operation must either be the
initial value $0$ or the value  
introduced by a write event, thus satisfying Assumption~\ref{ass:causality}.

Memory systems occurring in practice also have the property of 
{\em data independence\/}, that is, control decisions are oblivious to the
data values.
A cache line carries along with the actual program data a
few state bits for recording whether it is
in shared, exclusive or invalid mode.
Typically, actions do not depend on the value of the data in the cache
line. 
This can be observed, for example, in the memory system shown in
Figure~\ref{fig:example}. 
Note that there are no predicates involving the data fields of the
cache lines and the messages in any of the internal events of the system.
In such systems, renaming the data values of a run results in yet
another run of the system.
Moreover, every run can be obtained by data value renaming from some
run in which the initial value and values of write events to any
location $i$ are all distinct from each other.
In order to define data independence formally,
we define below the notion of an unambiguous run and the notion
of data value renaming.

Formally, a run $\sigma$ of $\msystem(n,m,v)$ is {\em unambiguous\/} if for
all~$1 \leq i \leq m$ and $x \in L^w(\sigma,i)$, we have 
(1)~$\data(\sigma(x)) \neq 0$, and 
(2)~$\data(\sigma(x)) \neq \data(\sigma(y))$ 
    for all $y \in L^w(\sigma,i)\setminus\{x\}$.
In an unambiguous run, every write event to a location $i$ has a value
distinct from the initial value of $i$ and the value of every other
write to $i$.
The trace $\trace{\sigma}$ corresponding to an unambiguous run $\sigma$ 
is called an {\em unambiguous trace}.
If a run is both unambiguous and causal, each read event to location
$i$ with data value $0$ reads the initial value of $i$, and each read
event with a nonzero data value reads the value written by the unique
write event with a matching data value.
Thus, a read event can be paired with its source write event
just by comparing data values.


A function $\lambda: \nat_m \times \natzero \rightarrow \natzero$ is
called a {\em renaming function\/} if $\lambda(j,0) = 0$ 
for all $1 \leq j \leq m$.
Intuitively, the function $\lambda$ provides for each memory location $c$ and
data value $d$ the renamed data value $\lambda(c,d)$.
Since $0$ models the fixed initial value of all locations, the function $\lambda$ 
does not rename the value $0$.
Let $\lambda^d$ be a function on $\memevent(n,m,v)$ such that 
for all $e=\zug{a,b,c,d} \in \memevent(n,m)$, we have 
$\lambda^d(e) = \zug{a,b,c,\lambda(c,d)}$.
The function $\lambda^d$ is extended to sequences in
$\memevent(n,m,v)^*$ in the natural way.

We state the data independence assumption formally as follows.
\begin{assumption}[Data independence]
\label{ass:data-independence}
For all $n,m,v \geq 1$ and sequences $\tau \in \memevent(n,m,v)^*$, 
we have that $\tau$ is a trace of $\msystem(n,m,v)$ iff 
there is an unambiguous trace $\tau'$ of $\msystem(n,m)$ and a renaming function 
$\lambda: \nat_m \times \natzero \rightarrow \natzero_v$ such that 
$\tau = \lambda^d(\tau')$. 
\end{assumption}

Assumptions~\ref{ass:causality} and~\ref{ass:data-independence}
are motivated by the data handling in typical cache-coherence
protocols.  
We can have these assumptions be true on protocol descriptions
by imposing restrictions on the operations allowed on variables 
that contain data values \cite{Nalumasu99}.
For example, one restriction can be that no data variable appears in the
guard expression of an internal event or in the control expression of a 
conditonal.

\section{Sequential consistency}
\label{sec:mmodel}
Suppose $\msystem(n,m,v)$ is a memory system for some $n,m,v \geq 1$.
The sequential consistency memory model \cite{Lamport79}
is a correctness requirement on the runs of $\msystem(n,m,v)$.
In this section, we define sequential consistency formally.

We first define the simpler notion of a sequence being serial.
For all $\tau \in \memevent(n,m,v)^*$ and
$1 \leq i \leq |\tau|$, let $\upto(\tau,i)$ be the set
$\{1 \leq k \leq i ~|~ \op(\tau(k)) = \wrt \wedge \loc(\tau(k)) =
\loc(\tau(i))\}$.
In other words, the set $\upto(\tau,i)$ is the set of write events in
$\tau$ to location $\loc(\tau(i))$ occurring not later than $i$.
A sequence $\tau \in \memevent(n,m,v)^*$ is 
{\em serial\/} if for all $1 \leq u \leq |\tau|$, we have
\[\begin{array}{ll}
\data(\tau(u)) = 0, & \mathrm{if}~\upto(\tau,u) = \emptyset \\
\data(\tau(u)) = \data(\tau(\maximum{\upto(\tau,u)})), & \mathrm{if}~\upto(\tau,u)
\neq \emptyset. 
\end{array}\]
Thus, a sequence is serial if every read to a location $i$ returns
the value of the latest write to $i$ if one exists, and the initial
value $0$ otherwise.\footnote{The decision to
model the initial values of all locations by the value $0$ is 
implicit in our definition of a serial sequence.}

The {\em sequential consistency\/} memory model $\msc$ is a function
that maps every sequence of memory events~$\tau \in
\memevent(n,m,v)^*$ and processor~$1 \leq i 
\leq n$ to a total order $\msc(\tau,i)$ on $P(\tau,i)$
defined as follows:
for all $u,v \in P(\tau,i)$, we have 
$\zug{u,v} \in \procorder{\msc}{\tau}{i}$ iff $u < v$.
A sequence $\tau$ is {\em sequentially consistent\/}
if there is a permutation $f$ on 
$\nat_{|\tau|}$ such that the following conditions are satisfied.
\begin{enumerate}
\item[C1]
For all $1 \leq u,v \leq |\tau|$ and~$1 \leq i \leq n$, 
if $\zug{u,v} \in \procorder{\msc}{\tau}{i}$ 
then $f(u) < f(v)$.
\item[C2]
The sequence $\tau' = \tau_{f^{-1}(1)} \tau_{f^{-1}(2)} \ldots \tau_{f^{-1}(|\tau|)}$
is serial.
\end{enumerate}
Intuitively, the sequence $\tau'$ is a permutation of the sequence $\tau$
such that the event at index $u$ in $\tau$ is moved to index $f(u)$ 
in $\tau'$.
According to C1, this permutation must respect the total order
$\msc(\tau,i)$ for all $1 \leq i \leq n$. 
According to C2, the permuted sequence must be serial.
A run $\sigma \in \msystem(n,m,v)$ is sequentially
consistent if $\trace{\sigma}$ satisfies $\msc$.
The memory system $\msystem(n,m,v)$ is sequentially consistent
iff every run of $\msystem(n,m,v)$ is sequentially consistent.

The memory system in Figure~\ref{fig:example} is supposed to be
sequentially consistent. 
Here is an example of a sequentially consistent run $\sigma$ of that memory
system, the corresponding trace $\tau$ of $\sigma$, and the sequence
$\tau'$ obtained by permuting $\tau$.
\begin{center}
$\sigma$ = \parbox{1in}{$\zug{\EXCACK,1,1}$\\
                   $\zug{\UPDATE,1}$\\
                   $\zug{\wrt,1,1,1}$\\ 
                   $\zug{\rd,2,1,0}$\\
                   $\zug{\UPDATE,2}$\\
                   $\zug{\SHDACK,2,1}$\\
                   $\zug{\UPDATE,2}$\\
                   $\zug{\rd,2,1,1}$}
$\tau = \trace{\sigma}$ = \parbox{1in}{$\zug{\wrt,1,1,1}$\\
                            $\zug{\rd,2,1,0}$\\
                            $\zug{\rd,2,1,1}$} 
$\tau'$ = \parbox{1in}{$\zug{\rd,2,1,0}$\\
                      $\zug{\wrt,1,1,1}$\\                      
                      $\zug{\rd,2,1,1}$} 
\end{center}
Sequential consistency orders the event $\tau(2)$ before  
the event $\tau(3)$ at processor 2.
Let $f$ be the permutation on $\nat_3$ defined by
$f(1) = 2$, $f(2) = 1$, and $f(3) = 3$.
The sequence $\tau'$ is the permutation of $\tau$ under $f$.
It is easy to check that both conditions C1 and C2 mentioned above are 
satisfied. 

In order to prove that a run of a memory system is sequentially consistent, one needs to
provide a reordering of the memory events of the run. 
This reordering should be serial and should respect the total orders imposed by
sequential consistency at each processor.
Since the memory systems we consider in this paper are data
independent, we only need to show sequential consistency for the
unambiguous runs of the memory system.
This reduction is stated formally in the following theorem.
\begin{theorem}
\label{thm:red-to-unamb}
For all $n,m \geq 1$, 
every trace of $\msystem(n,m)$ is sequentially consistent iff
every unambiguous trace of $\msystem(n,m)$ is sequentially consistent.
\end{theorem}
\begin{proof}
The $\Rightarrow$ case is trivial.

($\Leftarrow$)
Let $\tau$ be a trace of $\msystem(n,m,v)$ for some $v \geq 1$.
From Assumption~\ref{ass:data-independence} there is an unambiguous trace
$\tau'$ of $\msystem(n,m)$ and a renaming function 
$\lambda:\nat_m \times \natzero \rightarrow \natzero_v$
such that $\tau = \lambda^d(\tau')$.
Since $\tau'$ is sequentially consistent, we know that conditions C1 and C2
are satisfied by $\tau'$.
It is not difficult to see that both conditions C1 and C2 are
satisfied by $\lambda^d(\tau')$ as well.
Therefore $\tau$ is sequentially consistent.
\end{proof}

\section{Witness}
\label{sec:witness}
Theorem~\ref{thm:red-to-unamb} allows us to prove that a memory system
$\msystem(n,m,v)$ is sequentially consistent by proving that all
unambiguous runs in $\msystem(n,m)$ is sequentially consistent.  
In this section, we reduce the problem of checking
sequential consistency on an unambiguous run to the problem of
detecting a cycle in a constraint graph.

Consider a memory system $\msystem(n,m,v)$ for some fixed $n,m,v \geq 1$.
A {\em witness\/} $\witness$ for $\msystem(n,m,v)$ maps every
trace $\tau$ of $\msystem(n,m,v)$ and location $1 \leq i \leq m$ to a total
order $\locorder{\witness}{\tau}{i}$
on the set of writes $L^w(\tau,i)$ to location $i$. 
If the trace~$\tau$ is unambiguous, the total order $\locorder{\witness}{\tau}{i}$
on the write events to location~$i$ can be extended to a partial 
order $\locorder{\expand{\witness}}{\tau}{i}$ on 
all memory events (including read events) to location~$i$. 
If a read event~$r$ reads the value written by the write event~$w$,
the partial order puts~$r$ {\em after\/} $w$ and all write events
preceding $w$, and {\em before\/} all write events succeeding~$w$.
Formally, for every unambiguous trace $\tau$ of $\msystem(n,m,v)$,
location~$1 \leq i \leq m$, and $x,y \in L(\tau,i)$, we have that
$\zug{x,y} \in \locorder{\expand{\witness}}{\tau}{i}$ iff
one of the following conditions holds. 
\begin{enumerate}
\item
$\data(\tau(x)) = \data(\tau(y))$, $\op(\tau(x)) = \wrt$, and 
$\op(\tau(y)) = \rd$.
\item
$\data(\tau(x)) = 0$ and $\data(\tau(y)) \neq 0$.  
\item
$\exists a,b \in L^w(\tau,i)$ such that 
    $\zug{a,b} \in \locorder{\witness}{\tau}{i}$,
    $\data(\tau(a)) = \data(\tau(x))$, and 
    $\data(\tau(b)) = \data(\tau(y))$.
\end{enumerate}

We now show that the relation $\locorder{\expand{\witness}}{\tau}{i}$
is a partial order.
First, we need the following lemma about $\locorder{\expand{\witness}}{\tau}{i}$. 
\begin{lemma}
\label{lemma:almost-ord}
For all unambiguous traces $\tau$ of $\msystem(n,m,v)$, 
locations~$1 \leq i \leq m$
and $r,s,t \in L(\tau,i)$, if $\zug{r,s} \in  
\locorder{\expand{\witness}}{\tau}{i}$,
then either $\zug{r,t} \in \locorder{\expand{\witness}}{\tau}{i}$ or 
$\zug{t,s} \in \locorder{\expand{\witness}}{\tau}{i}$.
\end{lemma}
\begin{proof}
Since $\zug{r,s} \in \locorder{\expand{\witness}}{\tau}{i}$,
either $\data(\tau(s)) \neq 0$ or there is a $x \in L^w(\tau,i)$ such
that $\data(\tau(s)) = \data(\tau(x))$.  
Since $\tau$ is an unambiguous trace, we have that $\data(\tau(x))
\neq 0$.
Therefore, we get that $\data(\tau(s)) \neq 0$ in both cases.
If $\data(\tau(t)) = 0$ we immediately get that 
$\zug{t,s} \in \locorder{\expand{\witness}}{\tau}{i}$.
So suppose $\data(\tau(t)) \neq 0$.
Since $\tau$ is unambiguous, there is $y \in L^w(\tau,i)$ such that
$\data(\tau(t)) = \data(\tau(y))$.
We have three cases from the definition of
$\zug{r,s} \in \locorder{\expand{\witness}}{\tau}{i}$.
\begin{enumerate}
\item
$\data(\tau(r)) = \data(\tau(s))$, $\op(\tau(r)) = \wrt$, and 
and $\op(\tau(s)) = \rd$.
Since $\witness$ is a total order on $L^w(\tau,i)$, either 
$\zug{r,y} \in \locorder{\witness}{\tau}{i}$ or 
$\zug{y,r} \in \locorder{\witness}{\tau}{i}$. 
In the first case, we have 
$\zug{r,t} \in \locorder{\expand{\witness}}{\tau}{i}$.
In the second case, we have 
$\zug{t,s} \in \locorder{\expand{\witness}}{\tau}{i}$.
\item
$\data(\tau(r)) = 0$ and $\data(\tau(s)) \neq 0$.  
We get that $\zug{r,t} \in \locorder{\expand{\witness}}{\tau}{i}$.
\item
$\exists a,b \in L^w(\tau,i)$ such that 
    $\zug{a,b} \in \locorder{\witness}{\tau}{i}$,
    $\data(\tau(a)) = \data(\tau(r))$, and 
    $\data(\tau(b)) = \data(\tau(s))$.
Since $\witness$ is a total order on $L^w(\tau,i)$, either 
$\zug{a,y} \in \locorder{\witness}{\tau}{i}$ or 
$\zug{y,a} \in \locorder{\witness}{\tau}{i}$. 
In the first case, we have 
$\zug{r,t} \in \locorder{\expand{\witness}}{\tau}{i}$.
In the second case, we have by transitivity $\zug{y,b} \in
\locorder{\witness}{\tau}{i}$ and therefore 
$\zug{t,s} \in \locorder{\expand{\witness}}{\tau}{i}$.
\end{enumerate}
\end{proof}

\begin{lemma}
\label{lemma:partial-order}
For all unambiguous traces $\tau$ of $\msystem(n,m,v)$ and locations~$1 \leq i
\leq m$, we have that $\locorder{\expand{\witness}}{\tau}{i}$ is a
partial order.
\end{lemma}
\begin{proof}
We show that $\locorder{\expand{\witness}}{\tau}{i}$ is
irreflexive.
In other words, for all $1 \leq x \leq |\tau|$, we have that
$\zug{x,x} \not \in \locorder{\expand{\witness}}{\tau}{i}$.
This is an easy proof by contradiction by assuming 
$\zug{x,x} \in \locorder{\expand{\witness}}{\tau}{i}$ and
performing a case analysis over the three resulting
conditions. 

We show that $\locorder{\expand{\witness}}{\tau}{i}$ is
anti-symmetric. 
In other words, for all $1 \leq x < y \leq |\tau|$, if
$\zug{x,y} \in \locorder{\expand{\witness}}{\tau}{i}$ then
$\zug{y,x} \not \in \locorder{\expand{\witness}}{\tau}{i}$.
We do a proof by contradiction.
Suppose both $\zug{x,y} \in \locorder{\expand{\witness}}{\tau}{i}$ 
and $\zug{y,x} \in \locorder{\expand{\witness}}{\tau}{i}$.
We reason as in the proof of Lemma~\ref{lemma:almost-ord} to obtain
$\data(\tau(x)) \neq 0$ and $\data(\tau(y)) \neq 0$.
Therefore there are $a,b \in L^w(\tau,i)$ such that $\data(\tau(a)) =
\data(\tau(x))$ and $\data(\tau(b)) = \data(\tau(y))$.
We perform the following case analysis.
\begin{enumerate}
\item
$a = b$.
Either $\op(x) = \rd$ and $\op(y) = \rd$, 
or $\op(x) = \wrt$ and $\op(y) = \rd$,
or $\op(x) = \rd$ and $\op(y) = \wrt$.
In the first case $\zug{x,y} \not \in \locorder{\expand{\witness}}{\tau}{i}$ and
$\zug{y,x} \not \in \locorder{\expand{\witness}}{\tau}{i}$.
In the second case $\zug{y,x} \not \in \locorder{\expand{\witness}}{\tau}{i}$.
In the third case $\zug{x,y} \not \in \locorder{\expand{\witness}}{\tau}{i}$. 
\item
$\zug{a,b} \in \locorder{\witness}{\tau}{i}$.
We have $\data(\tau(x)) \neq \data(\tau(y))$ since $\tau$ is
unambiguous.
Since $\locorder{\witness}{\tau}{i}$ is a total order, we have 
$\zug{b,a} \not \in \locorder{\witness}{\tau}{i}$.
Therefore $\zug{y,x} \not \in \locorder{\expand{\witness}}{\tau}{i}$.
\item
$\zug{b,a} \in \locorder{\witness}{\tau}{i}$.
This case is symmetric to Case 2.
\end{enumerate}

Finally, we show that $\locorder{\expand{\witness}}{\tau}{i}$ is
transitive. 
Suppose $\zug{x,y} \in \locorder{\expand{\witness}}{\tau}{i}$ and 
$\zug{y,z} \in \locorder{\expand{\witness}}{\tau}{i}$.
From Lemma~\ref{lemma:almost-ord}, either 
$\zug{x,z} \in \locorder{\expand{\witness}}{\tau}{i}$ or
$\zug{z,y} \in \locorder{\expand{\witness}}{\tau}{i}$.
We have shown $\locorder{\expand{\witness}}{\tau}{i}$ to be
anti-symmetric.
Therefore $\zug{x,z} \in \locorder{\expand{\witness}}{\tau}{i}$.
\end{proof}

\subsection{Constraint graph}
Suppose $\tau$ is an unambiguous trace of $\msystem(n,m,v)$.
We have that $\procorder{\msc}{\tau}{i}$ is a total order on 
$P(\tau,i)$ for all~$1 \leq i \leq n$ from the definition of
sequential consistency.
We also have that $\locorder{\expand{\witness}}{\tau}{j}$ is a
partial order on $L(\tau,j)$ for all $1 \leq j \leq m$ from
Lemma~\ref{lemma:partial-order}. 
The union of the $n$ total orders $\procorder{\msc}{\tau}{i}$ and
$m$ partial orders $\locorder{\expand{\witness}}{\tau}{j}$ imposes a
graph on $\domain(\tau)$.
The acyclicity of this graph is a necessary and sufficient condition
for the trace $\tau$ to satisfy sequential consistency.
We define a function $G$ that for every 
witness $\witness$ returns a function $G(\witness)$. 
The function $G(\witness)$
maps every unambiguous trace $\tau$ of $\msystem(n,m,v)$ to the graph 
$\zug{\domain(\tau),
\bigcup_{1 \leq i \leq n} \procorder{\msc}{\tau}{i} \cup 
\bigcup_{1 \leq j \leq m} \locorder{\expand{\witness}}{\tau}{j}}$.
The work of Gibbons and Korach \cite{GibbonsKorach97} defines a constraint 
graph on the memory events of a run that is similar to $G(\witness)(\tau)$.

\begin{theorem}
\label{thm:witness-sat}
For all $n,m,v \geq 1$, every unambiguous trace of
$\msystem(n,m,v)$ is sequentially consistent iff 
there is a witness $\witness$ such that the graph $G(\witness)(\tau)$
is acyclic for every unambiguous trace $\tau$ of $\msystem(n,m,v)$. 
\end{theorem}
\begin{proof}
($\Rightarrow$) 
Suppose $\tau$ is an unambiguous trace of $\msystem(n,m,v)$. 
Then $\tau$ satisfies sequential consistency.
There is a permutation $f$ on $\nat_{|\tau|}$ such that conditions C1
and C2 are satisfied.
For all $1 \leq i \leq m$, define $\witness(\tau,i)$ to be the total
order on $L^w(\tau,i)$ such that for all $x,y \in L^w(\tau,i)$, 
we have $\zug{x,y} \in \witness(\tau,i)$ iff $f(x) < f(y)$.
We show that the permutation $f$ is a linearization of the vertices in 
$G(\witness)(\tau)$ that preserves all the edges.
In other words, if 
$\zug{x,y} \in \procorder{\msc}{\tau}{i}$ for some $1 \leq i \leq
n$ or $\zug{x,y} \in \locorder{\expand{\witness}}{\tau}{j}$ for some
$1 \leq j \leq m$, then $f(x) < f(y)$. 
If $\zug{x,y} \in \procorder{\msc}{\tau}{i}$ then we have from C1
that $f(x) < f(y)$. 
We show below that if 
$\zug{x,y} \in \locorder{\expand{\witness}}{\tau}{j}$ then 
$f(x) < f(y)$. 

Let $\tau' = \tau_{f^{-1}(1)} \tau_{f^{-1}(2)} \ldots \tau_{f^{-1}(|\tau|)}$.
For all $1 \leq u \leq |\tau|$ we have that $\tau(u) = \tau'(f(u))$.
We first show for all $a \in L^w(\tau,j)$ and $x \in L(\tau,j)$, if 
$\data(\tau(a)) = \data(\tau(x))$ then $f(a) \leq f(x)$.
Since $\tau$ is unambiguous, we have that 
$\data(\tau(a)) = \data(\tau(x)) \neq 0$.
Therefore $\data(\tau'(f(a))) = \data(\tau'(f(x))) \neq 0$.
We have that either $\op(\tau'(f(x))) = \rd$ or $x = a$.
In the first case $f(a) \in \upto(\tau',f(x))$ which implies
that $f(a) < f(x)$, and in the second case $f(a) = f(x)$.
Therefore $f(a) \leq f(x)$.

If $\zug{x,y} \in \locorder{\expand{\witness}}{\tau}{j}$ then we have
three cases.
In each case, we show that $f(x) < f(y)$. 
\begin{enumerate}
\item
$\data(\tau(x)) = \data(\tau(y))$, $\op(\tau(x)) = \wrt$, and 
$\op(\tau(y)) = \rd$.
Since $\tau$ is unambiguous $\data(\tau(x)) = \data(\tau(y)) \neq 0$.
We get that $\data(\tau'(f(y))) \neq 0$ which means that 
$\upto(\tau',f(y)) \neq \emptyset$ and $f(x) \in \upto(\tau',f(y))$.
Therefore $f(x) < f(y)$. 
\item
$\data(\tau(x)) = 0$ and $\data(\tau(y)) \neq 0$.  
Since $x \neq y$ we have $f(x) \neq f(y)$.
Suppose $f(y) < f(x)$.
Since $\data(\tau(y)) \neq 0$ there is $b \in L^w(\tau,j)$ such that 
$\data(\tau(b)) = \data(\tau(y))$.
Therefore we have that $f(b) \leq f(y) < f(x)$.
Therefore the set $\upto(\tau',f(x)) \neq \emptyset$.
Since $\tau'$ is unambiguous and $\data(\tau'(f(x))) = 0$
we have a contradiction.
\item
$\exists a,b \in L^w(\tau,j)$ such that 
    $\zug{a,b} \in \locorder{\witness}{\tau}{j}$,
    $\data(\tau(a)) = \data(\tau(x))$, and 
    $\data(\tau(b)) = \data(\tau(y))$.
We show $f(x) < f(y)$ by contradiction.
Suppose $f(x) = f(y)$. 
Then $x = y$ and $\data(\tau(a)) = \data(\tau(b))$.
Since $\tau$ is unambiguous we get $a = b$ which contradicts 
$\zug{a,b} \in \locorder{\witness}{\tau}{j}$.
Suppose $f(y) < f(x)$. 
We have that $f(a) \leq f(x)$ and
$f(b) \leq f(y)$.
Since $\zug{a,b} \in \locorder{\witness}{\tau}{j}$, we have $f(a)
< f(b)$ from the definition of $\witness$.
Thus we have $f(a) < f(b) \leq f(y) < f(x)$ 
Therefore $f(a) \neq \maximum{\upto(\tau',f(x))}$.
Since $\tau'$ is unambiguous and $\data(\tau'(f(a))) = 
\data(\tau'(f(x)))$ we have a contradiction.
\end{enumerate}

($\Leftarrow$)
Suppose $\tau$ is a trace of $\msystem(n,m,v)$.
Then there is a witness $\witness$ such that
$G(\witness)(\tau)$ is acyclic.
Let $f$ be a linearization of the vertices in
$G(\witness)(\tau)$ that respects all edges.
Then C1 is  satisfied.
Let $\tau'$ denote 
$\tau_{f^{-1}(1)} \tau_{f^{-1}(2)} \ldots \tau_{f^{-1}(|\tau|)}$.
Then we have that $\tau'(x) = \tau(f^{-1}(x))$ for all $1 \leq x \leq |\tau'|$.
For any $1 \leq x \leq |\tau'|$, suppose $\loc(\tau'(x)) = j$.
There are two cases.
\begin{enumerate}
\item
$\data(\tau'(x)) = 0$.
We show that $\upto(\tau',x) = \emptyset$.
Consider any vertex $1 \leq y \leq |\tau'|$ such that
$\op(\tau'(y)) = \wrt$ and $\loc(\tau'(y)) = j$.
Then $\tau(f^{-1}(x)) = 0$ and $\tau(f^{-1}(y)) \neq 0$.
Therefore $\zug{f^{-1}(x),f^{-1}(y)} \in \locorder{\expand{\witness}}{\tau}{j}$ and 
$\zug{f^{-1}(x),f^{-1}(y)}$ is an edge in $G(\witness)(\tau)$.
Therefore $f(f^{-1}(x)) < f(f^{-1}(y))$ or $x < y$.
Thus we have that $\upto(\tau',x) = \emptyset$.
\item
$\data(\tau'(x)) \neq 0$.
We show that $\upto(\tau',x) \neq \emptyset$ and 
if $y = \maximum{\upto(\tau',x)}$ then $\data(\tau'(x)) = \data(\tau'(y))$.
From Assumption~\ref{ass:causality}, there is $a \in
L^w(\tau',j)$ such that $\data(\tau'(a)) = \data(\tau'(x))$
and since $\tau'$ is unambiguous this write is unique.
Therefore $\data(\tau(f^{-1}(a))) = \data(\tau(f^{-1}(x)))$.
Either $f^{-1}(a) = f^{-1}(x)$ or $\op(\tau(f^{-1}(x))) = \rd$ in which case 
$\zug{f^{-1}(a),f^{-1}(x)} \in \locorder{\expand{\witness}}{\tau}{j}$.
In both cases, we have $a \leq x$ and therefore $\upto(\tau',x) \neq \emptyset$.
Consider any vertex $1 \leq b \leq |\tau'|$ such that 
$\op(\tau'(b)) = \wrt$, $\loc(\tau'(b)) = j$, and $a < b$.
Then $\zug{f^{-1}(a),f^{-1}(b)} \in \locorder{\witness}{\tau}{j}$ and
$\zug{f^{-1}(x),f^{-1}(b)} \in \locorder{\expand{\witness}}{\tau}{j}$.
Therefore $x < b$.
We thus get $a = \maximum{\upto(\tau',x)}$.
\end{enumerate}
\end{proof}

Theorems~\ref{thm:red-to-unamb} and~\ref{thm:witness-sat} can be combined easily
to yield the following theorem.
\begin{corollary}
\label{thm:unamb-witness-sat}
For all $n,m \geq 1$, every trace of
$\msystem(n,m)$ is sequentially consistent iff 
there is a witness $\witness$ such that the graph $G(\witness)(\tau)$
is acyclic for every unambiguous trace $\tau$ of $\msystem(n,m)$. 
\end{corollary}

\subsection{Simple witness}
\label{sec:simple-witness}
Corollary~\ref{thm:unamb-witness-sat} suggests that in order to prove
that the memory system $\msystem(n,m,v)$ is sequentially consistent,
we produce a witness $\witness$ and show for every unambiguous
trace $\tau$ of $\msystem(n,m)$ that the graph $G(\witness)(\tau)$ is
acyclic.
But the construction of the witness is still left to the verifier.
In this section, we argue that a simple witness, which orders the
write events to a location exactly in the order in which they occur,
suffices for a number of memory systems occurring in practice.
Formally, a witness $\witness$ is {\em simple\/} if for all traces $\tau$ of
$\msystem(n,m,v)$ and locations $1 \leq i \leq m$, we have
$\zug{x,y} \in \locorder{\witness}{\tau}{i}$ iff $x < y$
for all $x,y \in L^w(\tau,i)$.

Consider the memory system of Figure~\ref{fig:example}.
We argue informally that a simple witness is a good witness
for this memory system.
Permission to perform writes flows from one cache to another by means
of the $\EXCACK$ message.
Note that for each location $j$, the variable $\owner[j]$ is set to
$0$ (which is not the id of any processor) when an $\EXCACK$ message
is generated. 
When the $\EXCACK$ message is received at the destination (by the
$\UPDATE$ event), the destination moves to $\EXC$ state and sets
$\owner[j]$ to the destination id. 
A new $\EXCACK$ message is generated only when $\owner[j] \neq 0$.
Thus, the memory system has the property that 
each memory location can be held in $\EXC$ state by at most one cache.
Moreover, writes to the location $j$ can happen only when the cache has
the location in $\EXC$ state.
Therefore, at most one cache can be performing writes to a memory
location. 
This indicates that the logical order of the write events is the same
as their temporal order.
In other words, a simple witness is the correct witness for
demonstrating that a run is sequentially consistent.

In general, for any memory system in which at any time at most one
processor can perform write events to a location, a simple witness is
very likely to be the correct witness.
Most memory systems occurring in practice \cite{LLG90,KOH94,BDH99,BGM00} have
this property.   
In Section~\ref{sec:model-check}, we describe a model checking
algorithm to verify the correctness of a memory system with respect to
a simple witness.  
If a simple witness is indeed the desired witness and the memory
system is designed correctly, then our algorithm will be able to verify
its correctness.
Otherwise, it will produce an error trace suggesting to the verifier
that either there is an error in the memory system or the simple
witness is not the correct witness.
Thus our method for checking sequential consistency is clearly sound.
We have argued that it is also complete on most shared-memory systems that occur in
practice.

\section{Nice cycle reduction}
\label{sec:cycle}
For some $n,m,v \geq 1$, let $\msystem(n,m,v)$ be a memory system  
and $\witness$ a witness for it.
Let $\tau$ be an unambiguous trace of $\msystem(n,m,v)$.
Corollary~\ref{thm:unamb-witness-sat} tells us that the absence of cycles in
the graphs $G(\witness)(\tau)$ generated by the unambiguous traces of
$\msystem(n,m)$ is a necessary and sufficient condition for every trace of
$\msystem(n,m)$ to be sequentially consistent.
In this section, we show that it suffices to detect a special class of
cycles called nice cycles.
In Section~\ref{sec:model-check}, we will show that detection of nice
cycles can be performed by model checking.

We fix some $k \geq 1$ and 
use the symbol $\oplus$ to denote addition over
the additive group with elements $\nat_k$ and identity element $k$.
A {\em $k$-nice\/} cycle in $G(\witness)(\tau)$ is a sequence 
$u_1,v_1,\ldots,u_k,v_k$ of distinct vertices in
$\nat_{|\tau|}$ such that the following conditions are true.
\begin{enumerate}
\item
For all $1 \leq x \leq k$, we have $\zug{u_x,v_x} \in
\procorder{\msc}{\tau}{i}$ 
for some $1 \leq i \leq n$ and $\zug{v_x,u_{x\oplus1}} \in 
\locorder{\expand{\witness}}{\tau}{j}$
for some $1 \leq j \leq m$.
\item
For all $1 \leq x < y \leq k$ and for all $1 \leq i,j \leq n$, 
if $\zug{u_x,v_x} \in \procorder{\msc}{\tau}{i}$ and 
$\zug{u_y,v_y} \in \procorder{\msc}{\tau}{j}$ then $i \neq j$.
\item
For all $1 \leq x < y \leq k$ and for all $1 \leq i,j \leq m$, 
if $\zug{v_x,u_{x\oplus1}} \in \locorder{\expand{\witness}}{\tau}{i}$ and 
$\zug{v_y,u_{y\oplus1}} \in \locorder{\expand{\witness}}{\tau}{j}$ then $i \neq j$.
\end{enumerate}
In a $k$-nice cycle, no two edges belong to the relation 
$\procorder{\msc}{\tau}{i}$ for any processor $i$.
Similarly, no two edges belong to the relation 
$\locorder{\expand{\witness}}{\tau}{j}$ for any location $j$.
The above definition also implies that if a cycle is $k$-nice then 
$k \leq \minimum{\{n,m\}}$.

\begin{theorem}
\label{thm:cycle-nice}
If the graph $G(\witness)(\tau)$ has a cycle, then it has a
$k$-nice cycle for some $k$ such that $1 \leq k \leq \minimum{\{n,m\}}$.
\end{theorem}
\begin{proof}
Suppose $G(\witness)(\tau)$ has no $k$-nice cycles 
but does have a cycle.
Consider the shortest such cycle $u_1,\ldots,u_l$ where 
$l \geq 1$.
For this proof, we denote by $\oplus$ addition over
the additive group with elements $\nat_l$ and identity element $l$.
Then for all $1 \leq x \leq l$ either 
$\zug{u_x,u_{x\oplus1}} \in \procorder{\msc}{\tau}{i}$ for some $1
\leq i \leq n$ or  
$\zug{u_x,u_{x\oplus1}} \in \locorder{\expand{\witness}}{\tau}{i}$ 
for some $1 \leq i \leq m$.

Since the cycle $u_1,\ldots,u_l$ is not $k$-nice for any $k$, there are 
$1 \leq a < b \leq l$ such that either 
(1)~$\zug{u_a,u_{a\oplus1}} \in \procorder{\msc}{\tau}{i}$ 
and $\zug{u_b,u_{b\oplus1}} \in \procorder{\msc}{\tau}{i}$ 
for some $1 \leq i \leq n$, or 
(2)~$\zug{u_a,u_{a\oplus1}} \in \locorder{\expand{\witness}}{\tau}{i}$ 
and $\zug{u_b,u_{b\oplus1}} \in \locorder{\expand{\witness}}{\tau}{i}$ 
for some $1 \leq i \leq m$.

Case (1). 
We have from the definition of $\msc$ that $u_a < u_{a\oplus1}$ and 
$u_b < u_{b\oplus1}$.
Either $u_a < u_b$ or $u_b < u_a$.
If $u_a < u_b$ then $u_a < u_{b\oplus1}$ or
$\zug{u_a,u_{b\oplus1}} \in \procorder{\msc}{\tau}{i}$.
If $u_b < u_a$ then $u_b < u_{a\oplus1}$ or
$\zug{u_b,u_{a\oplus1}} \in \procorder{\msc}{\tau}{i}$.
In both cases, we have a contradiction since the cycle can be 
made shorter.

Case (2).
From Lemma~\ref{lemma:almost-ord}, either 
$\zug{u_a,u_b} \in \locorder{\expand{\witness}}{\tau}{i}$ or 
$\zug{u_b,u_{a\oplus1}} \in \locorder{\expand{\witness}}{\tau}{i}$.  
In both cases, we have a contradiction since the cycle can be made
shorter. 
\end{proof}

\section{Symmetry reduction}
\label{sec:symmetry}
Suppose $\msystem(n,m,v)$ is a memory system for some $n,m,v \geq 1$. 
In this section, we use symmetry arguments to further reduce the class
of cycles that need to be detected in constraint graphs.
Each $k$-nice cycle has $2 \times k$ edges with one edge each for $k$
different processors and $k$ different locations.
These edges can potentially occur in any order 
yielding a set of isomorphic cycles.
But if the memory system $\msystem(n,m,v)$
is symmetric with respect to processor and memory location ids,
presence of any one of  
the isomorphic nice cycles implies the existence of a nice
cycle in which the edges are arranged in a canonical order.
Thus, it suffices to search for a cycle with edges in a
canonical order.


We discuss processor symmetry in Section~\ref{sec:proc-sym} and
location symmetry in Section~\ref{sec:loc-sym}. 
We combine processor and location symmetry to demonstrate the
reduction from nice cycles to canonical nice cycles in
Section~\ref{sec:comb}. 

\subsection{Processor symmetry}
\label{sec:proc-sym}
For any permutation $\lambda$ on $\nat_n$, the function
$\lambda^p$ on $\memevent(n,m,v)$ permutes the processor ids of events
according to $\lambda$.
Formally, for all $e=\zug{a,b,c,d} \in \memevent(n,m,v)$, we define 
$\lambda^p(e) = \zug{a,\lambda(b),c,d}$.
The function $\lambda^p$ is extended to sequences in
$\memevent(n,m,v)^*$ in the natural way.

\begin{assumption}[Processor symmetry]
\label{ass:msystem-proc-sym}
For every permutation $\lambda$ on $\nat_n$ and for all traces $\tau$
of the memory system $\msystem(n,m,v)$, we have that $\lambda^p(\tau)$
is a trace of $\msystem(n,m,v)$.
\end{assumption}

We argue informally that the memory system in Figure~\ref{fig:example}
satisfies Assumption~\ref{ass:msystem-proc-sym}.
The operations performed by the various parameterized actions on
the state variables that store processor ids are symmetric.  
Suppose $s$ is a state of the system.
We denote by $\lambda^p(s)$ the state obtained by permuting
the values of variables that store processors ids according to $\lambda$.
Then, for example, if the action $\UPDATE(i)$ in some state $s$ yields state
$t$, then the action $\UPDATE(\lambda(i))$ in state $\lambda^p(s)$
yields the state $\lambda^p(t)$.
Thus, from any run $\sigma$ we can construct another run
$\lambda^p(\sigma)$.
If a shared-memory system is described with symmetric types, such as
scalarsets \cite{IpDill96}, used to model variables containing
processor ids, then it has the property of processor symmetry by
construction. 

The following lemma states that the sequential consistency memory
model is symmetric with respect to processor ids.
It states that two events in a trace $\tau$ ordered by sequential
consistency remain ordered under any permutation of processor ids.
\begin{lemma}
\label{lemma:mmodel-proc-sym}
Suppose $\lambda$ is a permutation on $\nat_n$.
Suppose $\tau$ and $\tau'$ are traces of $\msystem(n,m,v)$ such that 
$\tau' = \lambda^p(\tau)$.
Then for all $1 \leq x,y \leq |\tau|$, and for all $1 \leq i \leq n$,
we have that $\zug{x,y} \in \procorder{\msc}{\tau}{i}$ iff 
$\zug{x,y} \in \procorder{\msc}{\tau'}{\lambda(i)}$.
\end{lemma}
\begin{proof}
For all $1 \leq x,y \leq |\tau|$ and for all $1 \leq i \leq n$, we have that
\[\begin{array}{ll}
& \zug{x,y} \in \procorder{\msc}{\tau}{i} \\
\Leftrightarrow 
& \proc(\tau(x)) = \proc(\tau(y)) = i~\mathrm{and}~x < y \\
\Leftrightarrow 
& \proc(\tau'(x)) = \proc(\tau'(y)) = \lambda(i)~\mathrm{and}~x < y \\
\Leftrightarrow 
& \zug{x,y} \in \procorder{\msc}{\tau'}{\lambda(i)}.
\end{array}\]
\end{proof}

The following lemma states that the partial order $\expand{\witness}$ 
obtained from a simple witness $\witness$ is 
symmetric with respect to processor ids.
It states that two events to location $i$ 
ordered by $\locorder{\expand{\witness}}{\tau}{i}$ 
in a trace $\tau$ remain 
ordered under any permutation of processor ids.
\begin{lemma}
\label{lemma:expand-witness-proc-sym}
Suppose $\witness$ is a simple witness for the memory system
$\msystem(n,m,v)$ and $\lambda$ is a permutation on $\nat_n$.
Suppose $\tau$ and $\tau'$ are unambiguous traces of $\msystem(n,m,v)$
such that $\tau' = \lambda^p(\tau)$.
Then for all $1 \leq x,y \leq |\tau|$ and for all $1 \leq i \leq m$,
we have that $\zug{x,y} \in \locorder{\expand{\witness}}{\tau}{i}$ iff 
$\zug{x,y} \in \locorder{\expand{\witness}}{\tau'}{i}$.
\end{lemma}
\begin{proof}
We have $\zug{x,y} \in \locorder{\witness}{\tau}{i}$ iff $x < y$ iff
$\zug{x,y} \in \locorder{\witness}{\tau'}{i}$.
From the definition of $\locorder{\expand{\witness}}{\tau}{i}$ we have the 
following three cases.
\begin{enumerate}
\item
$\data(\tau(x)) = \data(\tau(y))$, $\op(\tau(x)) = \wrt$, 
$\op(\tau(y)) = \rd$ iff
$\data(\tau'(x)) = \data(\tau'(y))$, $\op(\tau'(x)) = \wrt$, 
$\op(\tau'(y)) = \rd$.
\item
$\data(\tau(x)) = 0$ and $\data(\tau(y)) \neq 0$ iff 
$\data(\tau'(x)) = 0$ and $\data(\tau'(y)) \neq 0$.
\item
$\exists a,b \in L^w(\tau,i)$ such that 
    $a < b$,
    $\data(\tau(a)) = \data(\tau(x))$, 
    $\data(\tau(b)) = \data(\tau(y))$ iff
$\exists a,b \in L^w(\tau',i)$ such that 
    $a < b$,
    $\data(\tau'(a)) = \data(\tau'(x))$, 
    $\data(\tau'(b)) = \data(\tau'(y))$.
\end{enumerate}
\end{proof}

\subsection{Location symmetry}
\label{sec:loc-sym}
For any permutation $\lambda$ on $\nat_m$, the function
$\lambda^l$ on $\memevent(n,m,v)$ permutes the location ids of events
according to $\lambda$.
Formally, for all $e=\zug{a,b,c,d} \in \memevent(n,m,v)$, we define 
$\lambda^l(e) = \zug{a,b,\lambda(c),d}$.
The function $\lambda^l$ is extended to sequences in
$\memevent(n,m,v)^*$ in the natural way.

\begin{assumption}[Location symmetry]
\label{ass:msystem-loc-sym}
For every permutation $\lambda$ on $\nat_m$ and for all traces 
$\tau$ of the memory system $\msystem(n,m,v)$, 
we have that $\lambda^l(\tau)$ is a trace of $\msystem(n,m,v)$.
\end{assumption}

We can argue informally that the memory system in
Figure~\ref{fig:example} satisfies
Assumption~\ref{ass:msystem-loc-sym} also.
The operations performed by the various parameterized actions on
the state variables that store location ids are symmetric.  
Suppose $s$ is a state of the system.
We denote by $\lambda^l(s)$ the state obtained by permuting
the values of variables that store location ids according to $\lambda$.
Then, for example, if the action $\UPDATE(i)$ in some state $s$ yields state
$t$, then the action $\UPDATE(\lambda(i))$ in state $\lambda^l(s)$
yields the state $\lambda^l(t)$.
If scalarsets are used for modeling variables containing location ids,
the shared-memory system will have the property of location symmetry
by construction. 

The following lemma states that the sequential consistency memory
model is symmetric with respect to location ids.
It states that two events in a trace $\tau$ ordered by sequential
consistency remain ordered under any permutation of location ids.

\begin{lemma}
\label{lemma:mmodel-loc-sym}
Suppose $\lambda$ is a permutation on $\nat_m$.
Suppose $\tau$ and $\tau'$ are traces of $\msystem(n,m,v)$ such that 
$\tau' = \lambda^l(\tau)$.
Then for all $1 \leq x,y \leq |\tau|$, and for all $1 \leq i \leq n$,
we have that $\zug{x,y} \in \procorder{\msc}{\tau}{i}$ iff 
$\zug{x,y} \in \procorder{\msc}{\tau'}{i}$.
\end{lemma}
\begin{proof}
For all $1 \leq x,y \leq |\tau|$ and for all $1 \leq i \leq m$, we have that
\[\begin{array}{ll}
& \zug{x,y} \in \procorder{\msc}{\tau}{i} \\
\Leftrightarrow 
& \proc(\tau(x)) = \proc(\tau(y)) = i~\mathrm{and}~x < y \\
\Leftrightarrow 
& \proc(\tau'(x)) = \proc(\tau'(y)) = i~\mathrm{and}~x < y \\
\Leftrightarrow 
& \zug{x,y} \in \procorder{\msc}{\tau'}{i}.
\end{array}\]
\end{proof}

The following lemma states that the partial order $\expand{\witness}$ 
obtained from a simple witness $\witness$ is 
symmetric with respect to location ids.
It states that two events to location $i$ 
ordered by $\locorder{\expand{\witness}}{\tau}{i}$ 
in a trace $\tau$ remain 
ordered under any permutation of location ids.

\begin{lemma}
\label{lemma:expand-witness-loc-sym}
Suppose $\witness$ is a simple witness for the memory system
$\msystem(n,m,v)$ and $\lambda$ is a permutation on $\nat_m$.
Suppose $\tau$ and $\tau'$ are unambiguous traces of $\msystem(n,m,v)$ such that 
$\tau' = \lambda^l(\tau)$.
Then for all $1 \leq x,y \leq |\tau|$ and for all $1 \leq i \leq m$,
we have that $\zug{x,y} \in \locorder{\expand{\witness}}{\tau}{i}$ iff 
$\zug{x,y} \in \locorder{\expand{\witness}}{\tau'}{\lambda(i)}$.
\end{lemma}
\begin{proof}
We have $\zug{x,y} \in \locorder{\witness}{\tau}{i}$ iff $x < y$ iff
$\zug{x,y} \in \locorder{\witness}{\tau'}{\lambda(i)}$.
From the definition of $\locorder{\expand{\witness}}{\tau}{i}$ we have the 
following three cases.
\begin{enumerate}
\item
$\data(\tau(x)) = \data(\tau(y))$, $\op(\tau(x)) = \wrt$, 
$\op(\tau(y)) = \rd$ iff 
$\data(\tau'(x)) = \data(\tau'(y))$, $\op(\tau'(x)) = \wrt$, 
$\op(\tau'(y)) = \rd$.
\item
$\data(\tau(x)) = 0$ and $\data(\tau(y)) \neq 0$ iff 
$\data(\tau'(x)) = 0$ and $\data(\tau'(y)) \neq 0$.
\item
$\exists a,b \in L^w(\tau,i)$ where 
    $a < b$,
    $\data(\tau(a)) = \data(\tau(x))$, and
    $\data(\tau(b)) = \data(\tau(y))$ iff 
$\exists a,b \in L^w(\tau',\lambda(i))$ where 
    $a < b$,
    $\data(\tau'(a)) = \data(\tau'(x))$, and
    $\data(\tau'(b)) = \data(\tau'(y))$.
\end{enumerate}
\end{proof}

\subsection{Combining processor and location symmetry}
\label{sec:comb}
We fix some $k \geq 1$ and 
use the symbol $\oplus$ to denote addition over
the additive group with elements $\nat_k$ and identity element $k$.
A $k$-nice cycle $u_1,v_1,\ldots,u_k,v_k$
is {\em canonical\/} if 
$\zug{u_x,v_x} \in \procorder{\msc}{\tau}{x}$ and 
$\zug{v_x,u_{x\oplus1}}~\in~\locorder{\expand{\witness}}{\tau}{x\oplus1}$
for all $1 \leq x \leq k$.
In other words, the processor edges in a canonical nice cycle are
arranged in increasing order of processor ids.
Similarly, the location edges are arranged in increasing order of
location ids.
The following theorem claims that if the constraint graph of a run has
a nice cycle then there is some run with a canonical nice cycle as well.

\begin{theorem}
\label{thm:cycle-canonical}
Suppose $\witness$ is a simple witness for the memory system
$\msystem(n,m,v)$. 
Let $\tau$ be an unambiguous trace of $\msystem(n,m,v)$.
If the graph $G(\witness)(\tau)$ has a $k$-nice cycle, then
there is an unambiguous trace $\tau''$ of $\msystem(n,m,v)$ such that
$G(\witness)(\tau'')$ has a canonical $k$-nice cycle.
\end{theorem}
\begin{proof}
Let $u_1,v_1,\ldots,u_k,v_k$ be a
$k$-nice cycle in $G(\witness)(\tau)$.
Let $1 \leq i_1,\ldots,i_k \leq n$ and $1 \leq j_1,\ldots,j_k
\leq m$ be such that $\zug{u_x,v_x} \in \procorder{\msc}{\tau}{i_x}$ and 
$\zug{v_x,u_{x\oplus1}} \in \locorder{\expand{\witness}}{\tau}{j_{x\oplus1}}$ 
for all $1 \leq x \leq k$.
Let $\alpha$ be a permutation on $\nat_n$ that
maps $i_x$ to $x$ for all $1 \leq x \leq k$.
Then from Assumption~\ref{ass:msystem-proc-sym} there is a trace
$\tau'$ of $\msystem(n,m,v)$ such that $\tau' = \alpha^p(\tau)$. 
Let $\beta$ be a permutation on $\nat_m$ that maps $j_x$ to $x$
for all $1 \leq x \leq k$.
Then from Assumption~\ref{ass:msystem-loc-sym} there is a trace $\tau''$
of $\msystem(n,m,v)$ such that $\tau'' = \beta^l(\tau')$. 
For all $1 \leq x \leq k$, we have that
\[\begin{array}{lll}
& \zug{u_x,v_x} \in \procorder{\msc}{\tau}{i_x} & \\
\Leftrightarrow
& \zug{u_x,v_x} \in \procorder{\msc}{\tau'}{\alpha(i_x)} =
\procorder{\msc}{\tau'}{x} 
& \mathrm{from~Lemma~\ref{lemma:mmodel-proc-sym}} \\
\Leftrightarrow
& \zug{u_x,v_x} \in \procorder{\msc}{\tau''}{x} 
& \mathrm{from~Lemma~\ref{lemma:mmodel-loc-sym}}.
\end{array}\]
For all $1 \leq x \leq k$, we also have that
\[\begin{array}{lll}
& \zug{v_x,u_{x\oplus1}} \in \locorder{\expand{\witness}}{\tau}{j_{x\oplus1}} & \\
\Leftrightarrow 
& \zug{v_x,u_{x\oplus1}} \in \locorder{\expand{\witness}}{\tau'}{j_{x\oplus1}} 
& \mathrm{from~Lemma~\ref{lemma:expand-witness-proc-sym}} \\
\Leftrightarrow 
& \zug{v_x,u_{x\oplus1}} \in
\locorder{\expand{\witness}}{\tau''}{\beta(j_{x\oplus1})} 
= \locorder{\expand{\witness}}{\tau''}{x\oplus1}
& \mathrm{from~Lemma~\ref{lemma:expand-witness-loc-sym}}.
\end{array}\]
Therefore $u_1,v_1,\ldots,u_k,v_k$ is a canonical
$k$-nice cycle in $G(\witness)(\tau'')$.
\end{proof}

Finally, Corollary~\ref{thm:unamb-witness-sat} and
Theorems~\ref{thm:cycle-nice} and~\ref{thm:cycle-canonical} yield the following 
theorem.
\begin{corollary}
\label{thm:nice-canonical-cycle-sat}
Suppose there is a simple
witness $\witness$ such that for all unambiguous traces $\tau$ of
$\msystem(n,m)$ 
the graph $G(\witness)(\tau)$ does not have a canonical
$k$-nice cycle for all $1 \leq k \leq \minimum{\{n,m\}}$.
Then every trace of $\msystem(n,m)$ is sequentially consistent.
\end{corollary}


\section{Model checking memory systems}
\label{sec:model-check}

\begin{figure}
\subfigure{
\parbox{2in}{
{\scriptsize
\begin{tabbing}
\hspace{0.2cm} \= xx \= xx \= xx \= xx \= xx \= xx \= \kill 
Automaton $\cons_k(j)$ for $1 \leq j \leq k$ \\
States $\{a,b\}$ \\
Initial state $a$ \\
Accepting states $\{a,b\}$ \\
Alphabet $\memevent(n,m,2)$ \\
Transitions \\
$[]~\neg (\op(e) = \wrt \wedge \loc(e) = j)$ \\
\>\> $\rightarrow s' = s$ \\ 
$[]~s = a \wedge \op(e) = \wrt \wedge \loc(e) = j \wedge \data(e) = 0$ \\
\>\> $\rightarrow s' = a$ \\ 
$[]~s = a \wedge \op(e) = \wrt \wedge \loc(e) = j \wedge \data(e) = 1$ \\
\>\> $\rightarrow s' = b$ \\ 
$[]~s = b \wedge \op(e) = \wrt \wedge \loc(e) = j \wedge \data(e) = 2$ \\
\>\> $\rightarrow s' = b$ 
\end{tabbing}
}
}
}\quad
\subfigure{
\parbox{2in}{
{\scriptsize
\begin{tabbing}
\hspace{0.2cm} \= xx \= xx \= xx \= xx \= xx \= xx \= \kill 
Automaton $\cons_k(j)$ for $k < j \leq m$ \\
States $\{a\}$ \\
Initial state $a$ \\
Accepting states $\{a\}$ \\
Alphabet $\memevent(n,m,2)$ \\
Transitions \\
$[]~\neg (\op(e) = \wrt \wedge \loc(e) = j) \vee \data(e) = 0$ \\
\>\> $\rightarrow s' = s$ 
\end{tabbing}
}
}
}
\caption{Automaton $\cons_k(j)$}
\label{fig:cons}
\end{figure}

\begin{figure}
{\scriptsize
\begin{tabbing}
\hspace{0.2cm} \= xx \= xx \= xx \= xx \= xx \= xx \= \kill 
Automaton $\checker_k(i)$ \\
States $\{a,b,\err\}$ \\
Initial state $a$ \\
Accepting states $\{\err\}$ \\
Alphabet $\memevent(n,m,2)$ \\
Transitions \\
$[]~s = a \wedge \proc(e) = i \wedge \loc(e) = i \wedge \data(e) \in \{1,2\}$ \\ 
\>\> $ \rightarrow s' = b$ \\
$[]~s = b \wedge \proc(e) = i \wedge \loc(e) = i\oplus1 \wedge 
(\data(e) = 0 \vee (\op(e) = \wrt \wedge \data(e) = 1))$ \\
\>\> $\rightarrow s' = \err$ \\
$[]~\mathit{otherwise}$ \\
\>\> $\rightarrow s' = s$
\end{tabbing}
}
\caption{Automaton $\checker_k(i)$}
\label{fig:automata-sc}
\end{figure}

Suppose $\msystem(n,m,v)$ is a memory system for some $n,m,v \geq 1$.
Let $\witness$ be a simple witness for $\msystem(n,m)$.
In this section, we present a model checking algorithm that, 
given a $k$ such that $1 \leq k \leq \minimum{\{n,m\}}$, 
determines whether there is a trace $\tau$ in $\msystem(n,m)$ such
that the graph $G(\witness)(\tau)$ has a canonical $k$-nice cycle.
Corollary~\ref{thm:nice-canonical-cycle-sat} then allows us to verify
sequential consistency on $\msystem(n,m,v)$ by
$\minimum{\{n,m\}}$ such model checking lemmas.
We fix some $k$ such that $1 \leq k \leq \minimum{\{n,m\}}$.
We use the symbol $\oplus$ to denote addition over
the additive group with elements $\nat_k$ and identity element $k$.
The model checking algorithm makes use of $m$ automata named $\cons_k(j)$ 
for $1 \leq j \leq m$, and $k$ automata named $\checker_k(i)$ for 
$1 \leq i \leq k$.
We define these automata formally below.


For all memory locations $1 \leq j \leq m$, let $\cons_k(j)$ be
the regular set of sequences in $\memevent(n,m,2)$ represented
by the automaton in Figure~\ref{fig:cons}. 
The automaton $\cons_k(j)$, when composed with $\msystem(n,m,v)$, 
constrains the write events to location $j$.
If $1 \leq j \leq k$ then $\cons_k(j)$ accepts traces where the first
few ($0$ or more) write events have data value $0$ followed by exactly one write
with data value $1$ followed by ($0$ or more) write events with data value $2$.
If $k < j \leq m$ then $\cons_k(j)$ accepts traces where all writes to
location $j$ have data value $0$.

For all $1 \leq i \leq k$, let $\checker_k(i)$ be the regular 
set of sequences in $\memevent(n,m,2)$ represented by the automaton in
Figure~\ref{fig:automata-sc}. 
The automaton $\checker_k(i)$ accepts a trace
$\tau$ if there are events $x$ and $y$ at processor $i$, with $x$
occurring before $y$, such that $x$ is an event to location $i$ with
data value $1$ or $2$ and $y$ is an event to location $i \oplus 1$
with data value $0$ or $1$.
Moreover, the event $y$ is required to be a write event if its data
value is $1$. 


\begin{figure}
\begin{center}
\input{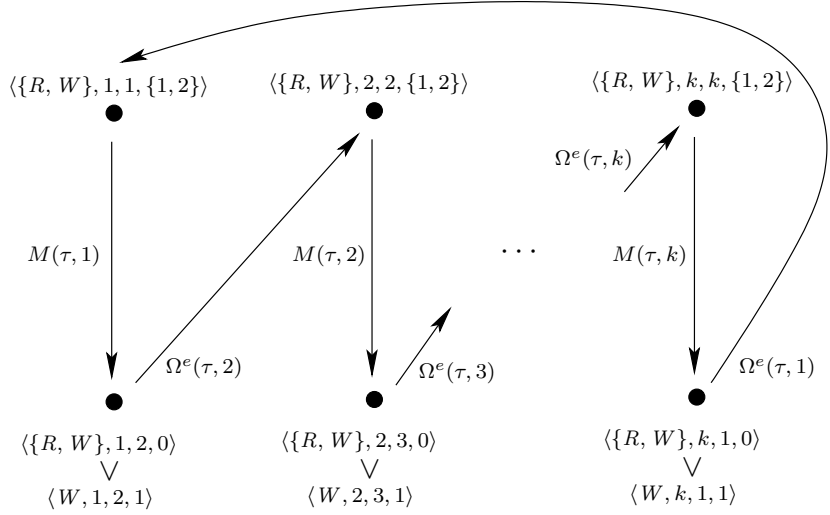}
\end{center}
\caption{Canonical $k$-nice cycle}
\label{fig:cycle}
\end{figure}

In order to check for canonical $k$-nice cycles, we compose the memory system
$\msystem(n,m,2)$ with $\cons_k(j)$ for all $1 \leq j \leq m$ and with 
$\checker_k(i)$ for all $1 \leq i \leq k$ and use a model checker to determine if
the resulting automaton has a run.  

Any accepting run of the composed system has $2 \times k$ events which
can be arranged as shown in Figure~\ref{fig:cycle} to yield a 
canonical $k$-nice cycle.
Each processor $i$ for $1 \leq i \leq k$ and 
each location $j$ for $1 \leq j \leq k$ supplies $2$ events.
Each event is marked by a $4$-tuple denoting the possible values for
that event.
For example, the $4$-tuple $\langle \{\rd,\wrt\},1,1,\{1,2\} \rangle$
denotes a read event or a write event by processor $1$ to location $1$ with
data value $1$ or $2$.
The edge labeled by $\msc(\tau,i)$ is due
to the total order imposed by sequential consistency on the events at processor $i$.   
The edge labeled by $\expand{\witness}(\tau,j)$ is due to the partial
order imposed by the simple witness on the events to location $j$.
For example, consider the edge labeled $\expand{\witness}(\tau,2)$
with the source event labeled by 
$\langle \{\rd,\wrt\},1,2,0 \rangle \vee \langle \wrt,1,2,1 \rangle$
and the sink event labeled by 
$\langle \{\rd,\wrt\},2,2,\{1,2\} \rangle$.
In any run of the composed system, the write events to location $2$ 
with value $0$ occur before the write event with value $1$ which
occurs before the write events with value $2$.
Since $\witness$ is a simple witness, 
the partial order $\expand{\witness}(\tau,2)$ 
orders all events labeled with $0$ before
all events labeled with $1$ or $2$.
Hence any event denoted by $\langle \{\rd,\wrt\},1,2,0 \rangle$ is
ordered before any event denoted by $\langle \{\rd,\wrt\},2,2,\{1,2\} \rangle$.  
Moreover, the unique write event to location $2$ with data value $1$
is ordered before any other events with value $1$ or $2$.
Hence the event $\langle \wrt,1,2,1 \rangle$ is ordered before 
any event denoted by $\langle \{\rd,\wrt\},2,2,\{1,2\} \rangle$.

We have given an intuitive argument above that a canonical $k$-nice
cycle can be constructed from any run in the composed system.
The following theorem proves that it is necessary and sufficient to
check that the composed system has a run.

\begin{theorem}
\label{thm:reduction}
There is a canonical
$k$-nice cycle in $G(\witness)(\tau)$ for some unambiguous trace $\tau$
of $\msystem(n,m)$
iff there is a trace $\tau'$ of $\msystem(n,m,2)$ such that 
$\tau' \in \cons_k(j)$ for all $1 \leq j \leq m$ and 
$\tau' \in \checker_k(i)$ for all $1 \leq i \leq k$.
\end{theorem}
\begin{proof}
$(\Rightarrow)$
Suppose there is a canonical $k$-nice cycle 
$u_1,v_1,\ldots,u_{k},v_{k}$ in the graph
$G(\witness)(\tau)$ for some unambiguous trace $\tau$ of $\msystem(n,m)$.  
Then $\zug{u_x,v_x} \in \procorder{\msc}{\tau}{x}$ and 
$\zug{v_x, u_{x\oplus1}} \in \locorder{\expand{\witness}}{\tau}{x\oplus1}$
for all $1 \leq x \leq k$.
From the definition of $\locorder{\expand{\witness}}{\tau}{x}$, we
have that $\data(\tau(x)) \neq 0$ for all $1 \leq x \leq k$.
Therefore, for all $1 \leq x \leq k$, there is a unique write event
$w_x$ such that $\data(\tau(w_x)) = \data(\tau(u_x))$.

For all $1 \leq j \leq m$, let $V_j$ be the set of data values written
by the write events to location $j$ in $\tau$, and let 
$f_j:V_j \rightarrow \nat_{|\tau|}$ be the function such that $f_j(v)$
is the index of the unique write event to location $j$ with data value $v$.
We define a renaming function $\lambda: \nat_m \times \natzero \rightarrow
\natzero_2$ as follows.
For all $k < j \leq m$ and $v \in \natzero$, 
we have $\lambda(j,x) = 0$.
For all $1 \leq j \leq k$ and $v \in \natzero$, we split the
definition into two cases.  
For $v \in V_j$, we have
\[
\begin{array}{lll}
\lambda(j,v) = & 0, & \mathrm{if}~f_j(v) < w_j \\
& 1, & \mathrm{if}~f_j(v) = w_j \\
& 2, & \mathrm{if}~f_j(v) > w_j.
\end{array}
\]
For $v \not \in V_j$, we have
\[
\begin{array}{lll}
\lambda(j,v) = & 0, & \mathrm{if}~v = 0 \\
& 2, & \mathrm{if}~v \neq 0.
\end{array}
\]
From Assumption~\ref{ass:data-independence},
there is a trace $\tau'$ of $\msystem(n,m,2)$ such that
$\tau' = \lambda^d(\tau)$. 
In $\tau'$, for every location $j$ such that $1 \leq j \leq k$ every 
write event before $w_j$ has the data value 0, the write event at $w_j$ 
has the data value 1, and the write events after $w_j$ have the data 
value $2$.
Moreover, for every location $j$ such that $k < j \leq m$ every write
event has the data value $0$.
Therefore $\tau' \in \cons_k(i)$ for all $1 \leq i \leq k$.

We show that $\tau' \in \checker_k(i)$ for all $1 \leq i \leq k$.
Since $\zug{u_i,v_i} \in \procorder{\msc}{\tau}{i}$,
we have that $u_i < v_i$ for all $1 \leq i \leq k$.
We already have that $\data(\tau'(u_i)) = \data(\tau'(w_i)) = 1$ for all $1 \leq i \leq k$.
Therefore all we need to show is that for all $1 \leq i \leq k$ we
have $\data(\tau'(v_i)) = 0$ or $\op(\tau'(v_i)) = \wrt$ and
$\data(\tau'(v_i)) = 1$. 
Since $\zug{v_i, u_{i\oplus1}} \in
\locorder{\expand{\witness}}{\tau}{i\oplus1}$, 
one of the following conditions hold.
\begin{enumerate}
\item
$\data(\tau(v_i)) = \data(\tau(u_{i\oplus1}))$, $\op(\tau(v_i)) = \wrt$, 
and $\op(\tau(u_{i\oplus1})) = \rd$.
We have that $\op(\tau'(v_i)) = \op(\tau(v_i)) = \wrt$.
Since $\data(\tau(v_i)) = \data(\tau(u_{i\oplus1}))$
we have $\data(\tau'(v_i)) = \data(\tau'(u_{i\oplus1})) = 1$.
Thus, we get $\op(\tau'(v_i)) = \wrt$ and 
$\data(\tau'(v_i)) = 1$.
\item
$\data(\tau(v_i)) = 0$ and $\data(\tau(u_{i\oplus1})) \neq 0$.  
From the definition of $\lambda$, we get that $\data(\tau'(v_i)) = 0$.
\item
$\exists a \in L^w(\tau,i\oplus1)$ such that 
    $\zug{a,w_{i \oplus 1}} \in \locorder{\witness}{\tau}{i\oplus1}$ and
    $\data(\tau(a)) = \data(\tau(v_i))$.
Since $\zug{a,b} \in \locorder{\witness}{\tau}{i\oplus1}$ and
$\witness$ is a simple witness we get $a < b$.
Therefore $\lambda(i \oplus 1, \data(\tau(a))) = 0$.
Thus $\lambda(i\oplus1,\data(\tau(v_i))) = 0$ and
$\data(\tau'(v_i)) = 0$.
\end{enumerate}
Thus, in all cases we have that either $\data(\tau'(v_i)) = 0$ or 
$\op(\tau'(v_i)) = \wrt$ and $\data(\tau'(v_i)) = 1$.
Therefore $\tau' \in \checker_k(i)$.

$(\Leftarrow)$
Suppose there is a trace $\tau'$ of $\msystem(n,m,2)$ such that
$\tau' \in \cons_k(j)$ for all $1 \leq j \leq m$ and
$\tau' \in \checker_k(i)$ for all $1 \leq i \leq k$.
For all $1 \leq i \leq k$, let $1 \leq u_i < v_i \leq |\tau'|$ be such
that the automaton $\checker_k(i)$ enters state $b$ for the first time
on observing $\tau'(u_i)$ and enters state $\err$ for the first time on
observing $\tau'(v_i)$.
Therefore we have $\proc(\tau'(u_i)) = i$, $\loc(\tau'(u_i)) = i$, and 
$\data(\tau'(u_i)) \in \{1,2\}$.
We also have $\proc(\tau'(v_i)) = i$, $\loc(\tau'(v_i)) = i\oplus1$, and
either $\data(\tau'(v_i)) = 0$ or $\op(\tau'(v_i)) = \wrt$ and
$\data(\tau'(v_i)) = 1$.
From Assumption~\ref{ass:data-independence}, there is an unambiguous
trace $\tau$ of $\msystem(n,m)$ and a renaming function $\lambda:\nat_m \times
\natzero \rightarrow \natzero_2$ such that $\tau' = \lambda^d(\tau)$.
We will show that $u_1,v_1,\ldots,u_k,v_k$ is a canonical $k$-nice
cycle in $G(\witness)(\tau)$.
Since $\proc(\tau(u_i)) = \proc(\tau(v_i)) = i$ and $u_i < v_i$, we
have $\zug{u_i,v_i} \in \procorder{\msc}{\tau}{i}$ for all $1 \leq
i \leq k$.
We show that 
$\zug{v_i, u_{i\oplus1}} \in 
\locorder{\expand{\witness}}{\tau}{i\oplus1}$
for all $1 \leq i \leq k$.
First $\loc(\tau(v_i)) = \loc(\tau(u_{i\oplus1})) = i\oplus1$.
For all $u,v \in L^w(\tau,i)$, if $\lambda(i,\data(\tau(u)) < 
\lambda(i,\data(\tau(v))$ then $u < v$ from the property of
$\cons_k(i)$. 
Since $\data(\tau'(u_{i\oplus1})) \in \{1,2\}$,
we have from the property of a renaming function that
$\data(\tau(u_{i\oplus1})) \neq 0$. 
There are two cases on $\tau'(v_i)$.
\begin{enumerate}
\item
$\data(\tau'(v_i)) = 0$.
There are two subcases: $\data(\tau(v_i)) = 0$ or 
$\lambda(i\oplus1,\data(\tau(v_i))) = 0$. 
In the first subcase, since $\data(\tau(u_{i\oplus1})) \neq 0$, we have 
$\zug{v_i, u_{i\oplus1}} \in \locorder{\expand{\witness}}{\tau}{i\oplus1}$.
In the second subcase, there are $a,b \in L^w(\tau,i\oplus1)$ such that
$\data(\tau(a)) = \data(\tau(v_i))$ and $\data(\tau(b)) =
\data(\tau(u_{i\oplus1}))$.
Since $\data(\tau'(a)) = 0$ and $\data(\tau'(b)) \in \{1,2\}$, we get from the
definition of $\cons_k(i\oplus1)$ that $a < b$ or 
$\zug{a,b} \in \locorder{\witness}{\tau}{i\oplus1}$.
Therefore $\zug{v_i, u_{i\oplus1}} \in 
\locorder{\expand{\witness}}{\tau}{i\oplus1}$.
\item
$\op(\tau'(v_i)) = \wrt$ and $\data(\tau'(v_i)) = 1$.
We have that $\op(\tau(v_i)) = \wrt$.
There is an event $b \in L^w(\tau,i\oplus1)$ such that
$\data(\tau(b)) = \data(\tau(u_{i\oplus1}))$.
There are two subcases: 
$\data(\tau'(u_{i\oplus1})) = 1$ or 
$\data(\tau'(u_{i\oplus1})) = 2$.
In the first subcase, we have $v_i=b$ since $\cons_k(i\oplus1)$ accepts
traces with a single write event labeled with 1.
Therefore $\data(\tau(v_i)) = \data(\tau(u_{i\oplus1}))$,
$\op(\tau(v_i)) = \wrt$ and $\op(\tau(u_{i\oplus1})) = \rd$, and we
get $\zug{v_i, u_{i\oplus1}} \in 
\locorder{\expand{\witness}}{\tau}{i\oplus1}$.
In the second subcase, since
$\data(\tau'(a)) = 1$ and $\data(\tau'(b)) = 2$, we get from the
definition of $\cons_k(i\oplus1)$ that $a < b$ or 
$\zug{a,b} \in \locorder{\witness}{\tau}{i\oplus1}$.
Therefore $\zug{v_i, u_{i\oplus1}} \in 
\locorder{\expand{\witness}}{\tau}{i\oplus1}$.
\end{enumerate}
Therefore $u_1,v_1,\ldots,u_k,v_k$ is a canonical $k$-nice
cycle in $G(\witness)(\tau)$.
\end{proof}


\begin{example}
We now give an example to illustrate the method described in this section.
Although the memory system in Figure~\ref{fig:example} is sequentially
consistent, an earlier version had an error.  
The assignment $\owner[j] := 0$ was missing in the guarded command of
the action $\zug{\SHDRSP,i,j}$.
We modeled the system in TLA+ \cite{Lamport94} and model checked the system
configuration with two processors and two locations using the model
checker TLC \cite{YML99}.
The error manifests itself while checking for the existence of a
canonical $2$-nice cycle.
The erroneous behavior is when the system starts in the initial
state with all cache lines in $\SHD$ state and $\owner[1] = \owner[2]
= 1$, and then executes the following sequence of 12 events:\\
$1.~\zug{\EXCACK,2,2}$\\
$2.~\zug{\UPDATE,2}$\\
$3.~\zug{\SHDACK,1,2}$\\
$4.~\zug{\EXCACK,2,2}$\\
$5.~\zug{\EXCACK,1,1}$\\
$6.~\zug{\UPDATE,1}$\\ 
$7.~\zug{\UPDATE,1}$\\
$8.~\zug{\wrt,1,1,1}$\\
$9.~\zug{\rd,1,2,0}$\\ 
$10.~\zug{\UPDATE,2}$\\
$11.~\zug{\wrt,2,2,1}$\\ 
$12.~\zug{\rd,2,1,0}$\\
After event~2, $\owner[2] = 2$, $\cache[1][2].s = \INV$, and
$\cache[2][2].s = \EXC$. 
Now processor~$1$ gets a shared ack message $\zug{\SHDACK,1,2}$ for
location~$2$.
Note that in the erroneous previous version of the example, this event
does not set $\owner[2]$ to~$0$.
Consequently $\owner[2] = 2$ and $\cache[2][2].s = \SHD$ after event~$3$.
An exclusive ack to processor~$2$ for location~$2$ is therefore
allowed to happen at event~$4$. 
Since the shared ack message to processor~$1$ in event~$3$ is still
sitting in $\inQ[1]$, $\cache[1][2].s$ is still $\INV$.
Therefore event~$4$ does not generate an $\INVAL$ message to
processor~$1$ for location~$2$.
At event~$5$, processor~$1$ gets an exclusive ack message for
location~$1$. 
This event also inserts an $\INVAL$ message on location~$1$ in $\inQ[2]$ 
behind the $\EXCRSP$ message on location~$2$.
After the $\UPDATE$ events to processor~$1$ in events~$6$ and~$7$, we
have $\cache[1][1].s = \EXC$ and $\cache[1][2].s = \SHD$.
Processor~$1$ writes~$1$ to location~$1$ and reads~$0$ from
location~$2$ in the next two events, thereby sending automaton
$\checker_2(1)$ to the state $\err$.
Processor~$2$ now processes the $\EXCRSP$ message to location~$2$ in
the $\UPDATE$ event~$10$. 
Note that processor~$2$ does not process the $\INVAL$ message to location~$1$ 
sitting in $\inQ[2]$.
At this point, we have $\cache[2][1].s = \SHD$ and $\cache[2][2].s = \EXC$.
Processor~$2$ writes~$1$ to location~$2$ and reads~$0$ from
location~$1$ in the next two events, thereby sending automaton
$\checker_2(2)$ to the state $\err$.
Since there has been only one write event of data value~$1$ to each
location, the run is accepted by $\cons_2(1)$ and $\cons_2(2)$ also.
\end{example}

Note that while checking for canonical $k$-nice cycles $\cons_k(j)$
has $2$ states for all $1 \leq j \leq k$ and $1$ state for $k < j \leq
m$.
Also $\checker_k(i)$ has $3$ states for all $1 \leq i \leq k$.
Therefore, by composing $\cons_k(j)$ and $\checker_k(i)$ with the
memory system $\msystem(n,m,2)$ we increase the state of the system by a factor of at
most $2^k \times 3^k$.
Actually, for all locations $k < j \leq m$ we are restricting write
events to have only the data value $1$.
Therefore, in practice we might reduce the set of reachable states.

\section{Related work}
\label{sec:rel-work}
Descriptions of shared-memory systems are parameterized by 
the number of processors, the number of memory locations, and the
number of data values.
The specification for such a system can
be either an invariant or a shared-memory model.
These specifications can be verified
for some fixed values of the parameters or for
arbitrary values of the parameters. 
The contribution of this paper is to provide a completely
automatic method based on model checking to verify the sequential
consistency memory model for fixed parameter values.
We now describe the related work on verification of shared-memory
systems along the two axes mentioned above.

A number of papers have looked at invariant verification.
Model checking has been used for fixed parameter values 
\cite{McMillanSchwalbe91,CGH93,EirikssonMcMillan95,IpDill96}, while
mechanical theorem proving \cite{LoewensteinDill92,ParkDill96} has
been used for arbitrary parameter values.
Methods combining automatic abstraction with model checking
\cite{PongDubois95,Delzanno00} have been used to verify snoopy
cache-coherence protocols for arbitrary parameter values.
McMillan \cite{McMillan01} has used a combination of
theorem proving and model checking to verify the directory-based FLASH
cache-coherence protocol \cite{KOH94} for arbitrary parameter values.
A limitation of all these approaches is that they do not explicate the
formal connection between the verified invariants and shared-memory
model for the protocol.

There are some papers that have looked at verification of shared-memory models.
Systematic manual proof methods \cite{LLOR99,PSCH98} and theorem
proving \cite{Arons01} have been used to verify sequential
consistency for arbitrary parameter values.  
These approaches require a significant amount of effort on the part of
the verifier.
Our method is completely automatic and is
a good debugging technique which can be applied before
using these methods.
The approach of Henzinger {\em et al.} \cite{HQR99b}
and Condon and Hu \cite{CondonHu01} requires a manually constructed
finite state machine called the serializer.
The serializer generates the witness total order for each run of the
protocol. 
By model checking the system composed of the protocol and the serializer,
it can be easily checked that the witness total order for every run is a
trace of serial memory.
This idea is a particular instance 
of the more general ``convenient computations''
approach of Katz and Peled \cite{KatzPeled92}.
In general, the manual construction of the serializer can be tedious and
infeasible in the case when unbounded storage is required.
Our work is an improvement since the witness total order is deduced 
automatically from the simple write order.
Moreover, the amount of state we add to the cache-coherence protocol
in order to perform the model checking is significantly less than that
added by the serializer approach.
The ``test model checking'' approach of Nalumasu {\em et al.}
\cite{NGMG98} can check a variety of memory models and is automatic.
Their tests are sound but incomplete for sequential consistency.
On the other hand, our 
method offers sound and complete verification for a large class of
cache-coherence protocols.

Recently Glusman and Katz \cite{GlusmanKatz01} have shown that, in
general, interpreting sequential consistency over finite traces is not
equivalent to interpreting it over infinite traces.
They have proposed conditions on shared-memory systems under which the
two are equivalent.
Their work is orthogonal to ours and a combination of the two will
allow verification of sequential consistency over infinite traces for
finite parameter values.

\section{Conclusions}
\label{sec:conclusions}
We now put the results of this paper in perspective.
Assumption~\ref{ass:causality} about causality
and Assumption~\ref{ass:data-independence} about data independence
are critical to our result that reduces the problem of verifying
sequential consistency to model checking.
Assumption~\ref{ass:msystem-proc-sym} about processor symmetry
and Assumption~\ref{ass:msystem-loc-sym} about location symmetry
are used to reduce the number of model checking lemmas to
$\minimum{\{n,m\}}$ rather than exponential in $n$ and $m$.

In this paper, the read and write events have been modeled as atomic events.
In most real machines, each read or write event is broken into two
separate events ---a request from the processor to the cache, and a 
response from the cache to the processor.   
Any memory model including sequential consistency naturally specifies 
a partial order on the requests.  
If the memory system services processor requests in order then the order of 
requests is the same as the order of responses. 
In this case, the method described in this paper can be used
by identifying the atomic read and write events with the responses.  
The case when the memory system services requests 
out of order is not handled by this paper.

The model checking algorithm described in the paper is sound and complete with 
respect to a simple witness for the memory system.  
In some protocols, for example the {\em lazy caching\/} protocol \cite{ABM89}, 
the correct witness is not simple.  
But the basic method described in the paper where data values of
writes are constrained by automata can still be used if
ordering decisions about writes can be made before the written values
are read. 
The lazy caching protocol has this property and extending the methods
described in the paper to handle it is part of our future work.
We would also like to extend our work to handle other memory
models.

\bibliographystyle{alpha}
\bibliography{/udir/qadeer/texinputs/diss}

\newcommand{\etalchar}[1]{$^{#1}$}
\begin{thebibliography}{NGMG98}

\bibitem[ABM93]{ABM89}
Y.~Afek, G.~Brown, and M.~Merritt.
\newblock Lazy caching.
\newblock {\em ACM Transactions on Programming Languages and Systems},
  15(1):182--205, 1993.

\bibitem[AMP96]{AMP96b}
R.~Alur, K.L. McMillan, and D.~Peled.
\newblock Model-checking of correctness conditions for concurrent objects.
\newblock In {\em Proceedings of the 11th Annual IEEE Symposium on Logic in
  Computer Science}, pages 219--228, 1996.

\bibitem[Aro01]{Arons01}
T.~Arons.
\newblock Using timestamping and history variables to verify sequential
  consistency.
\newblock In G.~Berry, H.~Comon, and A.~Finkel, editors, {\em CAV 01:
  Computer-aided Verification}, Lecture Notes in Computer Science 2102, pages
  423--435. Springer-Verlag, 2001.

\bibitem[BDH{\etalchar{+}}99]{BDH99}
E.~Bilir, R.~Dickson, Y.~Hu, M.~Plakal, D.~Sorin, M.~Hill, and D.~Wood.
\newblock Multicast snooping: {A} new coherence method using a multicast
  address network.
\newblock In {\em Proceedings of the 26th Annual International Symposium on
  Computer Architecture ({ISCA}'99)}, 1999.

\bibitem[BGM{\etalchar{+}}00]{BGM00}
L.A. Barroso, K.~Gharachorloo, R.~McNamara, A.~Nowatzyk, S.~Qadeer, B.~Sano,
  S.~Smith, R.~Stets, and B.~Verghese.
\newblock Piranha: a scalable architecture based on sigle-chip multiprocessing.
\newblock In {\em Proceedings of the 27st Annual International Symposium on
  Computer Architecture}, pages 282--293. IEEE Computer Society Press, 2000.

\bibitem[CE81]{ClarkeEmerson81}
E.M. Clarke and E.A. Emerson.
\newblock Design and synthesis of synchronization skeletons using
  branching-time temporal logic.
\newblock In {\em Workshop on Logic of Programs}, Lecture Notes in Computer
  Science 131, pages 52--71. Springer-Verlag, 1981.

\bibitem[CGH{\etalchar{+}}93]{CGH93}
E.M. Clarke, O.~Grumberg, H.~Hiraishi, S.~Jha, D.E. Long, K.L. McMillan, and
  L.A. Ness.
\newblock Verification of the {Futurebus+} cache coherence protocol.
\newblock In {\em Proceedings of the 11th IFIP WG10.2 International Conference
  on Computer Hardware Description Languages and their Applications}, pages
  15--30, 1993.

\bibitem[CH01]{CondonHu01}
A.E. Condon and A.J. Hu.
\newblock Automatable verification of sequential consistency.
\newblock In {\em 13th Symposium on Parallel Algorithms and Architectures}.
  ACM, 2001.

\bibitem[Com98]{alpha-manual}
Alpha~Architecture Committee.
\newblock {\em Alpha Architecture Reference Manual}.
\newblock Digital Press, 1998.

\bibitem[Del00]{Delzanno00}
G.~Delzanno.
\newblock Automatic verification of parameterized cache coherence protocols.
\newblock In E.A. Emerson and A.P. Sistla, editors, {\em CAV 2000: Computer
  Aided Verification}, Lecture Notes in Computer Science 1855, pages 53--68.
  Springer-Verlag, 2000.

\bibitem[EM95]{EirikssonMcMillan95}
{\'{A}}.Th. E{\'{\i}}riksson and K.L. McMillan.
\newblock Using formal verification/ $\!$analysis methods on the critical path in
  system design: a case study.
\newblock In P.~Wolper, editor, {\em CAV 95: Computer Aided Verification},
  Lecture Notes in Computer Science 939, pages 367--380. Springer-Verlag, 1995.

\bibitem[GK97]{GibbonsKorach97}
P.B. Gibbons and E.~Korach.
\newblock Testing shared memories.
\newblock {\em SIAM Journal on Computing}, 26(4):1208--1244, 1997.

\bibitem[GK01]{GlusmanKatz01}
M.~Glusman and S.~Katz.
\newblock Extending memory consistency of finite prefixes to infinite
  computations.
\newblock In K.G. Larsen and M.~Nielsen, editors, {\em CONCUR 01: Theories of
  Concurrency}, Lecture Notes in Computer Science, 
  Springer-Verlag, 2001.

\bibitem[HQR99]{HQR99b}
T.A. Henzinger, S.~Qadeer, and S.K. Rajamani.
\newblock Verifying sequential consistency on shared-memory multiprocessor
  systems.
\newblock In N.~Halbwachs and D.~Peled, editors, {\em CAV 99: Computer Aided
  Verification}, Lecture Notes in Computer Science 1633, pages 301--315.
  Springer-Verlag, 1999.

\bibitem[ID96]{IpDill96}
C.N. Ip and D.L. Dill.
\newblock Better verification through symmetry.
\newblock {\em Formal Methods in System Design}, 9(1--2):41--75, 1996.

\bibitem[KOH{\etalchar{+}}94]{KOH94}
J.~Kuskin, D.~Ofelt, M.~Heinrich, J.~Heinlein, R.~Simoni, K.~Gharachorloo,
  J.~Chapin, D.~Nakahira, J.~Baxter, M.~Horowitz, A.~Gupta, M.~Rosenblum, and
  J.~Hennessy.
\newblock The {S}tanford {FLASH} multiprocessor.
\newblock In {\em Proceedings of the 21st Annual International Symposium on
  Computer Architecture}, pages 302--313. IEEE Computer Society Press, 1994.

\bibitem[KP92]{KatzPeled92}
S.~Katz and D.~Peled.
\newblock Verification of distributed programs using representative
  interleaving sequences.
\newblock {\em Distributed Computing}, 6(2):107--120, 1992.

\bibitem[Lam78]{Lamport78}
L.~Lamport.
\newblock Time, clocks, and the ordering of events in a distributed program.
\newblock {\em Communications of the ACM}, 21(7):558--565, 1978.

\bibitem[Lam79]{Lamport79}
L.~Lamport.
\newblock How to make a multiprocessor computer that correctly executes
  multiprocess programs.
\newblock {\em IEEE Transactions on Computers}, C-28(9):690--691, 1979.

\bibitem[Lam94]{Lamport94}
L.~Lamport.
\newblock The {T}emporal {L}ogic of {A}ctions.
\newblock {\em ACM Transactions on Programming Languages and Systems},
  16(3):872--923, 1994.

\bibitem[LD92]{LoewensteinDill92}
P.~Loewenstein and D.L. Dill.
\newblock Verification of a multiprocessor cache protocol using simulation
  relations and higher-order logic.
\newblock {\em Formal Methods in System Design}, 1(4):355--383, 1992.

\bibitem[LLG{\etalchar{+}}90]{LLG90}
D.~Lenoski, J.~Laudon, K.~Gharachorloo, A.~Gupta, and J.~Hennessy.
\newblock The directory-based cache coherence protocol for the {DASH}
  multiprocessor.
\newblock In {\em Proceedings of the 17th Annual International Symposium on
  Computer Architecture}, pages 148--159, 1990.

\bibitem[LLOR99]{LLOR99}
P.~Ladkin, L.~Lamport, B.~Olivier, and D.~Roegel.
\newblock Lazy caching in {TLA}.
\newblock {\em Distributed Computing}, 12(2/3):151--174, 1999

\bibitem[McM01]{McMillan01}
K.L. McMillan.
\newblock Parameterized verification of the flash cache coherence protocol by
  compositional model checking.
\newblock In {\em CHARME 01: IFIP Working Conference on Correct Hardware Design
  and Verification Methods}, Lecture Notes in Computer Science, 
  Springer-Verlag, 2001.

\bibitem[MS91]{McMillanSchwalbe91}
K.L. McMillan and J.~Schwalbe.
\newblock Formal verification of the {Encore Gigamax} cache consistency
  protocol.
\newblock In {\em Proceedings of the International Symposium on Shared Memory
  Multiprocessors}, pages 242--251, 1991.

\bibitem[Nal99]{Nalumasu99}
R.P. Nalumasu.
\newblock {\em Formal Design and Verification Methods for Shared Memory
  Systems}.
\newblock PhD thesis, University of Utah, 1999.

\bibitem[NGMG98]{NGMG98}
R.P. Nalumasu, R.~Ghughal, A.~Mokkedem, and G.~Gopalakrishnan.
\newblock The `test model-checking' approach to the verification of formal
  memory models of multiprocessors.
\newblock In A.J. Hu and M.Y. Vardi, editors, {\em CAV 98: Computer Aided
  Verification}, Lecture Notes in Computer Science 1427, pages 464--476.
  Springer-Verlag, 1998.

\bibitem[PD95]{PongDubois95}
F.~Pong and M.~Dubois.
\newblock A new approach for the verification of cache coherence protocols.
\newblock {\em IEEE Transactions on Parallel and Distributed Systems},
  6(8):773--787, 1995.

\bibitem[PD96]{ParkDill96}
S.~Park and D.L. Dill.
\newblock Protocol verification by aggregation of distributed transactions.
\newblock In R.~Alur and T.A. Henzinger, editors, {\em CAV 96: Computer Aided
  Verification}, Lecture Notes in Computer Science 1102, pages 300--310.
  Springer-Verlag, 1996.

\bibitem[PSCH98]{PSCH98}
M.~Plakal, D.J. Sorin, A.E. Condon, and M.D. Hill.
\newblock Lamport clocks: verifying a directory cache-coherence protocol.
\newblock In {\em Proceedings of the 10th Annual ACM Symposium on Parallel
  Algorithms and Architectures}, pages 67--76, 1998.

\bibitem[QS81]{QueilleSifakis81}
J.~Queille and J.~Sifakis.
\newblock Specification and verification of concurrent systems in {CESAR}.
\newblock In M.~{Dezani-Ciancaglini} and U.~Montanari, editors, {\em Fifth
  International Symposium on Programming}, Lecture Notes in Computer Science
  137, pages 337--351. Springer-Verlag, 1981.

\bibitem[WG99]{sparc-manual}
D.L. Weaver and T.~Germond, editors.
\newblock {\em The {SPARC} Architecture Manual}.
\newblock Prentice Hall Inc., 1999.

\bibitem[YML99]{YML99}
Y.~Yu, P.~Manolios, and L.~Lamport.
\newblock Model checking {TLA+} specifications.
\newblock In {\em CHARME 99: IFIP Working Conference on Correct Hardware Design
  and Verification Methods}, Lecture Notes in Computer Science 1703, pages
  54--66. Springer-Verlag, 1999.

\end{thebibliography}

\end{document}